\documentclass[sigconf]{acmart}

\AtBeginDocument{%
  \providecommand\BibTeX{{%
    \normalfont B\kern-0.5em{\scshape i\kern-0.25em b}\kern-0.8em\TeX}}}

\setcopyright{acmcopyright}
\copyrightyear{2018}
\acmYear{2018}
\acmDOI{10.1145/1122445.1122456}

\acmConference[Woodstock '18]{Woodstock '18: ACM Symposium on Neural
  Gaze Detection}{June 03--05, 2018}{Woodstock, NY}
\acmBooktitle{Woodstock '18: ACM Symposium on Neural Gaze Detection,
  June 03--05, 2018, Woodstock, NY}
\acmPrice{15.00}
\acmISBN{978-1-4503-XXXX-X/18/06}

\usepackage{soul}
\usepackage{url}
\usepackage{epstopdf}
\usepackage{float}
\usepackage[utf8]{inputenc}
\usepackage{graphicx}
\usepackage{amsmath}
\usepackage{booktabs}
\usepackage[noend]{algorithmic}
\usepackage{algorithm}

\usepackage{booktabs} 
\usepackage{graphicx,array}
\usepackage{balance}
\usepackage{color,soul,xspace}

\usepackage{comment}
\usepackage{epsfig}
\usepackage{subfig}
\usepackage{mathrsfs,mdwlist,enumitem,amsthm,bm}

\setlist{nolistsep,leftmargin=*}

\newtheorem{definition}{\textbf{Definition}}

\begin{document}

\title{An Exploratory Study of Stock Price Movements from Earnings Calls}


\author{Sourav Medya$^1$, Mohammad Rasoolinejad$^1$, Yang Yang$^2$, Brian Uzzi$^1$}

\affiliation{%
  \institution{$^1$Kellogg School of Management \&  Northwestern Institute on Complex Systems, Northwestern University}
  \institution{$^2$Syracuse University}
}
\email{{sourav.medya,mohammad.rasoolinejad, uzzi}@kellogg.northwestern.edu}
\email{yyang87@syr.edu}

\renewcommand{\shortauthors}{Medya et al.}

\begin{abstract}
Financial market analysis has focused primarily on extracting signals from accounting, stock price, and other numerical “hard” data reported in P\&L statements or earnings per share reports.  Yet, it is well-known that decision-makers routinely use “soft” text-based documents that interpret the hard data they narrate.  Recent advances in computational methods for analyzing unstructured and soft text-based data at scale offer possibilities for understanding financial market behavior that could improve investments and market equity. A critical and ubiquitous form of soft data are earnings calls. Earnings calls are periodic (often quarterly) statements usually by CEOs who attempt to influence investors’ expectations of a company’s past and future performance. Here, we study the statistical relationship between earnings calls, company sales, stock performance, and analysts’ recommendations. Our study covers a decade of observations with approximately 100,000 transcripts of earnings calls from 6,300 public companies from January 2010 to December 2019. In this study, we report three novel findings. First, the buy, sell and hold recommendations from professional analysts made prior to the earnings have low correlation with stock price movements after the earnings call. Second, using our graph neural network based method that processes the semantic features of earnings calls, we reliably and accurately predict stock price movements in five major areas of the economy. Third, the semantic features of transcripts are more predictive of stock price movements than sales and earnings per share, i.e., traditional hard data in most of the cases. 

\end{abstract}


\maketitle

\section{Introduction}

Earnings calls are the discussion sessions that happen after the public companies release their financial data for the reporting period which is usually a quarter of the year or a fiscal year. Analysts,
investors, and journalists are often present during the earnings calls and the recordings of these can be accessed through corresponding
company websites.


In this session, usually the CEO and/or other representatives from the management of the company present their financial achievements during the last quarter and give guidelines for the next. The management usually discusses details about important company information such as growth, risks, purchases, liabilities, lawsuits, share buybacks, increase/decrease in dividends, any change in executive teams and future goals. The session usually consists of management discussion and analysis (MD\&A section) followed by a question-answering session involving the audience and the investors. 


The earnings call is a major event as the stock price market reflects higher level of volatility and trading volume prior, during and after earnings calls \cite{donders2000options}. Such volatility can result in bad investments, missing profit opportunities and huge losses for the investors. The large amount of available data on transcripts and stock market prices give us an opportunity to predict the directions of stock price movements more accurately. Moreover, there are other factors such as sentiment on social media and news that can also affect the stock market and have been studied in the literature \cite{xu2018stock,hu2018listening}. However, in this paper, we focus on associating the stock price movements from the transcripts of the earnings calls \cite{ma2020earnings,qin2019you}.

\begin{figure}[t]
	\vspace{-1mm}
	\centering
	\includegraphics[width=0.32\textwidth]{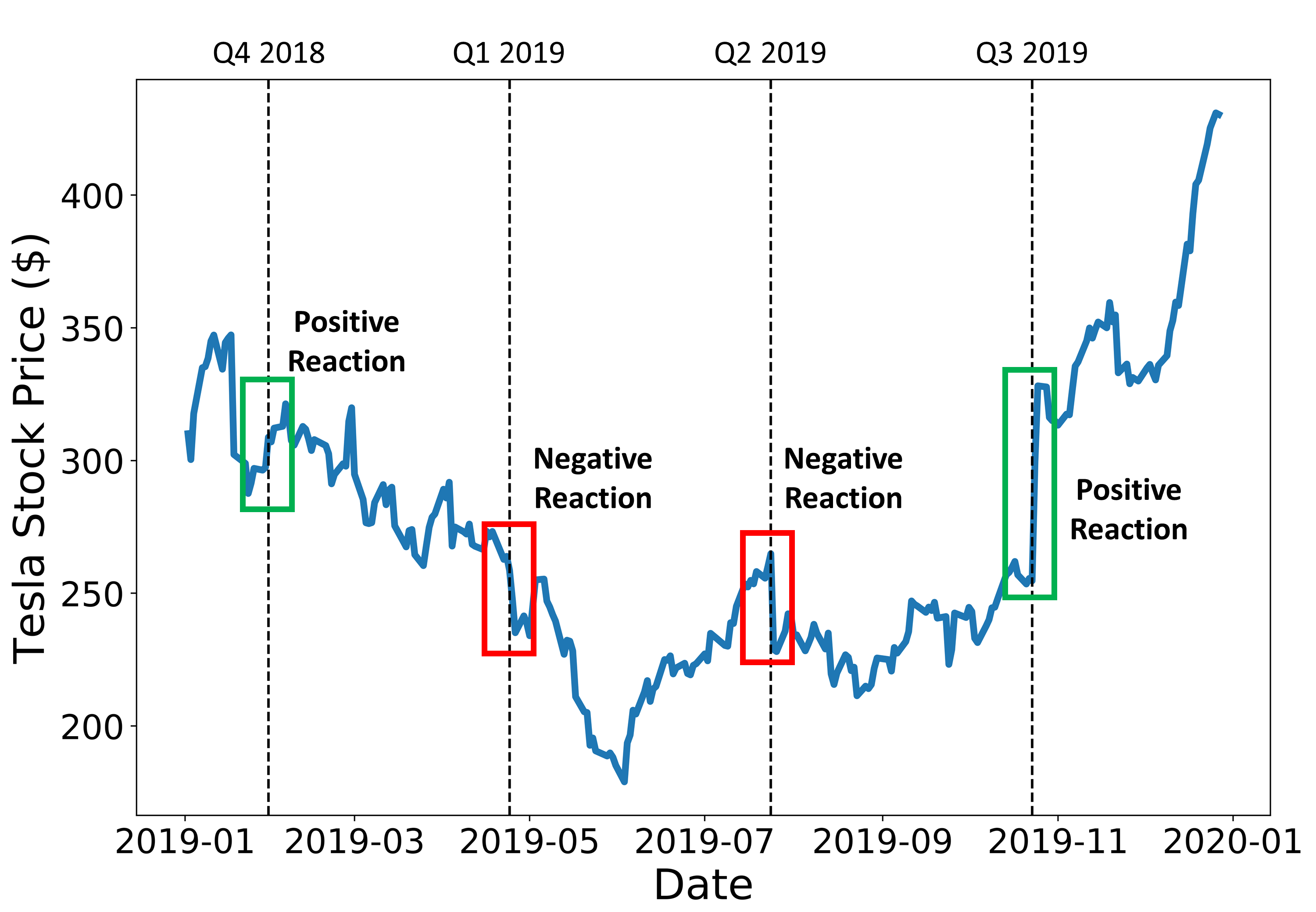}
			\vspace{-2mm}
	\caption{Reactions on stock price movements around occurrences of earnings calls from Tesla, Inc. during 2019. }

	\label{fig:reaction_tesla}
	\vspace{-2mm}
\end{figure}

Figure \ref{fig:reaction_tesla} shows an example of the stock price changes for the company Tesla, Inc. in 2019. The price movement after the occurrence of the earnings call can be positive or negative. As the figure shows daily effect, we define the label of the transcripts based on one day effect in the stock price (Def. \ref{defn:vbl} and Def. \ref{defn:sbl}). While the price change is important, it is also meaningful to compare the change with respect to the broader market behavior. To achieve that we compare the rate of the stock price changes with the corresponding index (Table \ref{tab:tab1}) over the next five business days. The labels of the transcripts are decided based on the majority of out-performances or under-performances with respect to the index value (Def. \ref{defn:ibl}).




In this paper, we perform an extensive study on the power of the earnings calls to predict stock price movements. We formally define the stock price movement as three different classification problems. First, we establish that the analysts' ratings prior to the earnings call are not indicative of the stock price movement prediction while the emotions in the transcripts play a significant role. This motivates us to design more rigorous methods based on semantics of the transcripts. Using a graph neural network based method, we show that the earnings calls can predict upward or downward movements of the stock prices more accurately than alternative signals. Furthermore, while the differences between actual earnings and sales with the estimated ones impact the stock price movements, these have lower predictive power than the earnings call. 

Our main contributions can be summarized as follows:
\vspace{-1mm}
\begin{itemize}
\item We study the problem of stock movement prediction from earnings calls and formulate three classification problems considering the daily and weekly effect of the earnings call on stock prices.

\item We provide descriptive analyses on labels as well as sentiments for our created large dataset of approximately 100,000 transcripts.

\item From the recommendations of analysts, we demonstrate that analysts' ratings available prior to the earnings calls do not have a significant impact on the stock price changes while the emotion traits available in the transcripts play a significant role in predicting stock price movements.

\item Through experimental evaluation, we show that our proposed graph neural network based method can predict stock price movement from earnings call transcripts with relatively high quality.

\item We show that the transcripts have more predictive power than actual and estimated values of sales and earnings per share.  In particular, the semantic features of transcripts produce at least 10\% higher average recall and 33\% higher average precision in two of the fastest growing areas of the economy, ``Technology" and ``Services".

\end{itemize}


\section{Stock Price Movements}
\label{sec:problem_definition}
We formulate the stock price movements as binary labels. Let $\mathbb{C} = \{C_1,C_2,\cdots, C_m\}$ be the set of $m$ companies. We denote the closing stock price of the company $c$ on the day $d$ as $S_d^c$. As the stock prices vary during a day, we consistently use the closing price of the stock on a particular day throughout our analysis. For a company $c$, the set of $t$ transcripts are denoted by $\mathbb{T}^c=\{T^c_{d_1},T^c_{d_2},\cdots, T^c_{d_t}\}$ where $T^c_{d_i}$ represents the earnings call transcript on day $d_i$. The value of $t$ can vary depending on the history of earnings calls.

\subsection{Daily Movement}
\label{sec:dsmp}
Our aim is to measure the local impact of the earnings call transcripts on the corresponding stock values. Thus, we define stock price movements (positive or negative) based on the stock values on the day before and after the occurrence of the earnings call. More specifically, for a company $c$, if the earnings call happen on the day $d$, the stock values on the $d-1$ and $d+1$ are compared and an upward and downward movements get binary values of 1 and 0 respectively. The $d+1$ and $d-1$ days denote the next and previous business days of the day $d$. Formally, this value based stock movement label of a transcript can be defined as follows:

\begin{definition}
 \label{defn:vbl}
\textbf{Value Based Label Function (VBL):} We define the label function $y_v(T^c_d)$ $\in \{0,1\}$ for a transcript $T^c_d$ of a company $c$ on the day $d$ as follows :

\begin{equation*}
    y_v(T^c_d) =
    \begin{cases}
        1, & \text{if $S^c_{d+1}>S^c_{d-1}$}\\
        0, & \text{otherwise}
    \end{cases}
\end{equation*}
where $d+1$ and $d-1$ denote the next and previous business days respectively of the day $d$.
\end{definition}

The previous definition does not account for the value of the change. We capture a significant amount of change (\textbf{shock}) both in upward and downward directions in another label function:
\begin{definition}
 \label{defn:sbl}
\textbf{Shock Based Label Function (SBL):} We define the label function $y_s(T^c_d)$ $\in \{0,1\}$ for a transcript $T^c_d$ of a company $c$ on the day $d$ as follows :

\begin{equation*}
    y_s(T^c_d) =
    \begin{cases}
        1, & \text{if $\frac{S^c_{d+1}-S^c_{d-1}}{S^c_{d-1}}\geq \tau$}\\
        0, & \text{if $\frac{S^c_{d-1}-S^c_{d+1}}{S^c_{d-1}}\geq \tau$}
    \end{cases}
\end{equation*}
where $d+1$ and $d-1$ denote the next and previous business days respectively of the day $d$; $\tau$ is a threshold (we set it as $.05$ i.e., $5\%$).
\end{definition}

Our aim is to investigate the predictive power of the earnings call transcripts of the stock price movement. To do so, we learn a prediction function $f$, where the features constructed from the earning calls are given as input and the defined labels ($y_v$ and $y_s$) act as output variables (Sec. \ref{sec:method}).

\begin{figure*}[ht]
\vspace{-4mm}
    \centering
    \captionsetup[subfigure]{labelfont={normalsize,bf},textfont={normalsize,bf}}
    \subfloat[Positive ]{\includegraphics[width=0.24\textwidth]{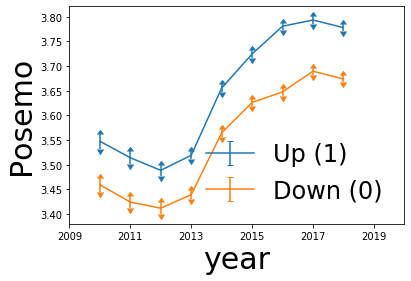}\label{fig:neg_five}}
    \subfloat[Negative ]{\includegraphics[width=0.24\textwidth]{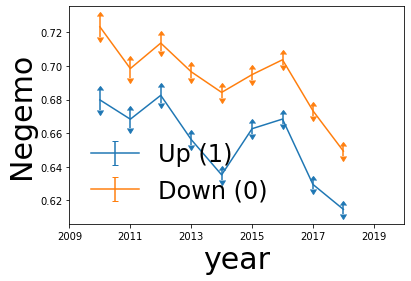} \label{fig:pos_five}}
    \subfloat[Sad ]{\includegraphics[width=0.24\textwidth]{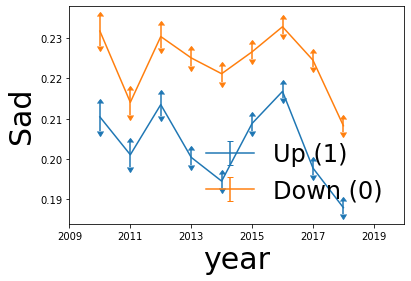} \label{fig:pos_five}}
    \subfloat[Anxiety ]{\includegraphics[width=0.24\textwidth]{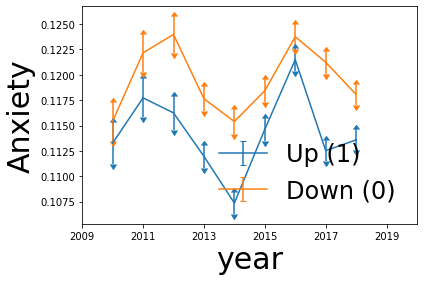} \label{fig:pos_five}}
    \vspace{-2mm}
    \caption{\textbf{Value Based Label ($y_{v}$): }The average sentiment scores in the transcripts over the years for both labels ($0$ and $1$). As expected, while the positive sentiment is higher for upward labels over the years; negative, sad and anxiety sentiments are higher for transcripts downward labels.} 
\label{fig:sentiment_daily_years}
\end{figure*}

\begin{figure*}[ht]
\vspace{-7mm}
    \centering
    \captionsetup[subfigure]{labelfont={normalsize,bf},textfont={normalsize,bf}}
    \subfloat[Positive ]{\includegraphics[width=0.24\textwidth]{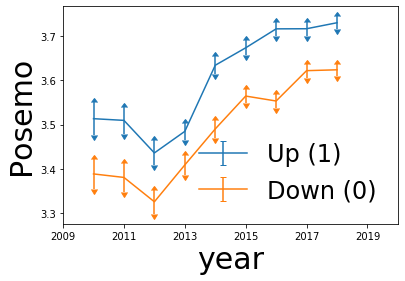}\label{fig:neg_five}}
    \subfloat[Negative ]{\includegraphics[width=0.24\textwidth]{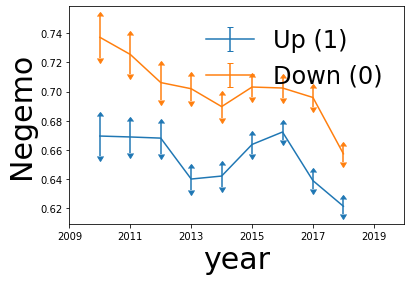} \label{fig:pos_five}}
    \subfloat[Sad ]{\includegraphics[width=0.24\textwidth]{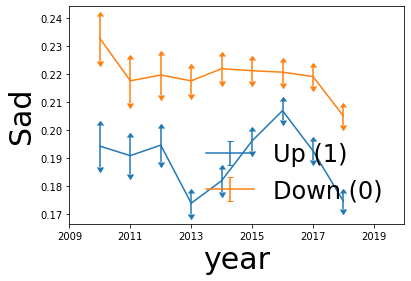} \label{fig:pos_five}}
    \subfloat[Anxiety ]{\includegraphics[width=0.24\textwidth]{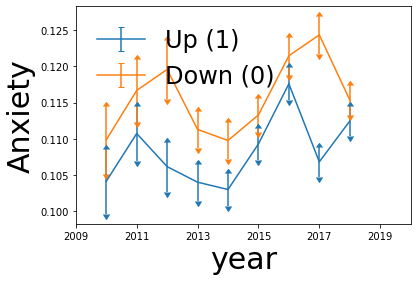} \label{fig:pos_five}}

    \vspace{-2mm}
    \caption{\textbf{Shock Based Label ($y_{s}$): }The average sentiment scores in the transcripts over the years for both labels ($0$ and $1$). As expected, consistent with the case in value based label, the positive sentiment  higher for transcripts with upward labels; negative, sad and anxiety sentiments are higher for transcripts with downward labels.} 
\label{fig:sentiment_daily_years_shock}
\end{figure*}

\textbf{Sector-wise comparison:} In the experiments (Sec. \ref{sec:acc_results}), we perform classification tasks based on the aforementioned labels for the companies that are in the same sector. Each index is a measure showing the performance of companies in terms of stock values in the corresponding sector. The indices for different sectors are given in Table \ref{tab:tab1}.  Note that we compute the index value based labels using the corresponding sector index.

\subsection{Weekly \& Normalized Movement}
The previous definitions of the labels for transcripts do not capture the stock movement compared to the broader market such as companies in the same sector. We further use index (see Table \ref{tab:tab1}) values to make a comparison and define positive and negative movements accordingly. Additionally, the previous labels consider stock values for only one day after the earnings call happens. We extend this notion of locality to one week (i.e., five business days) and compare it with the index values for next five business days.

Let $I_d$ denote the index value for the day $d$. We compare the rates of increase in the stock prices and index values. Let an earnings call ($T^c_d$) occur for a company $c$ on the day $d$. The rate of increase in the value is computed based on the previous day. So, the rate of increase ($R$) on the day $d+1$ for the stock value would be, $R(S^c_{d+1}) = \frac{S^c_{d+1}-S^c_{d}}{S^c_{d}}$. The stock movement is upward on the day $d+1$ if $R(S^c_{d+1})> R(I_{d+1})$ and we denote this upward movement by $1$. Otherwise, we call it downward movement and denote by $0$. From this definition we construct a $5$-dimensional vector $\mathbb{L}(T^c_d)$ containing binary values for the next five business days after the day $d$. We define a parameter $k$ that controls the output of label based on $k$ zeros and ones in $\mathbb{L}(T^c_d)$.
\begin{definition} \label{defn:ibl}
\textbf{Index Based Label Function (IBL):} We define the label function $y_{I,k}(T^c_d)$ $\in \{0,1\}$ for a transcript $T^c_d$ of a company $c$ on the day $d$ as follows :

\begin{equation*}
    y_{I,k}(T^c_d) =
    \begin{cases}
        1, & \text{if $\mathbb{L}(T^c_d)$ has $\geq k$ 1s}\\
        0, & \text{if $\mathbb{L}(T^c_d)$ has $\geq k$ 0s}
    \end{cases}
\end{equation*}
\end{definition}

Similarly, as in Sec. \ref{sec:dsmp}, we learn a prediction function where the features constructed from the earning calls are given as input and the defined label ($y_{I,k}$) acts as an output variable. Note that we consider $k$ to be between $3$ and $5$ as we define the final label based on majority of zeros or ones in $\mathbb{L}$. 

While the daily (value and shock) labels are relevant, they are miopoic in terms of investments and have high implied volatility \cite{donders2000options}. The weekly index based label (IBL) reflects a company’s performance over a week (less miopic and less volatile). This also helps to show whether the earnings calls are a predictor of the stock price movement for a longer duration.



\begin{table}[ht]
	\centering
	\vspace{-2mm}
	\begin{tabular}{ |c|c|c| }
		\hline
		Sector & Ref. Index & Percentage \\
		\hline
		Services (Service) & IYC & 18.5 \\
		Technology (Tech) & XLK & 18.3 \\
		Financial (Fin) & XLF & 18.1 \\
		Healthcare(Health) & XLV & 12.6 \\
		Basic Materials (Mat) & XLB & 11.3 \\
		Consumer (Con) & XLY & 9.2 \\
		Industrial (Ind) & XLI & 8.8 \\
		Utilities (Util) & XLU & 3.0 \\
		Not Specified & SP500 & 0.78 \\
		\hline
	\end{tabular}
	\caption{Percentage of transcripts in different company sectors: the sector of a company is determined from the reference index (Ref. Index).} \label{tab:tab1}
	\vspace{-4mm}
\end{table}


\begin{figure*}[ht]
\vspace{-7mm}
    \centering
    \captionsetup[subfigure]{labelfont={normalsize,bf},textfont={normalsize,bf}}
    
    \subfloat[ Year-wise for $y_v$]{\includegraphics[width=0.22\textwidth]{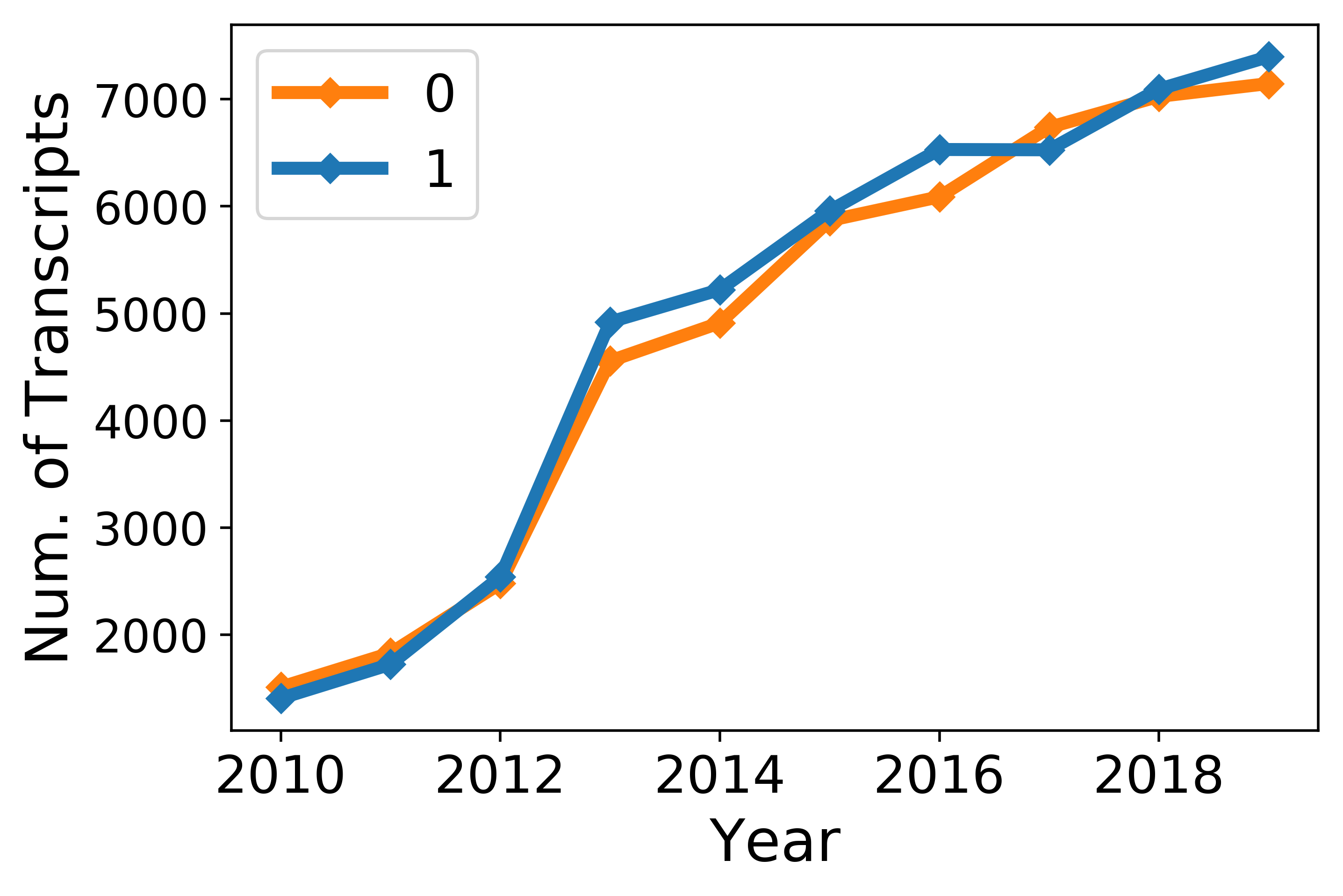}\label{fig:data_year_daily}}
    \subfloat[Quarter-wise for $y_v$ ]{\includegraphics[width=0.22\textwidth]{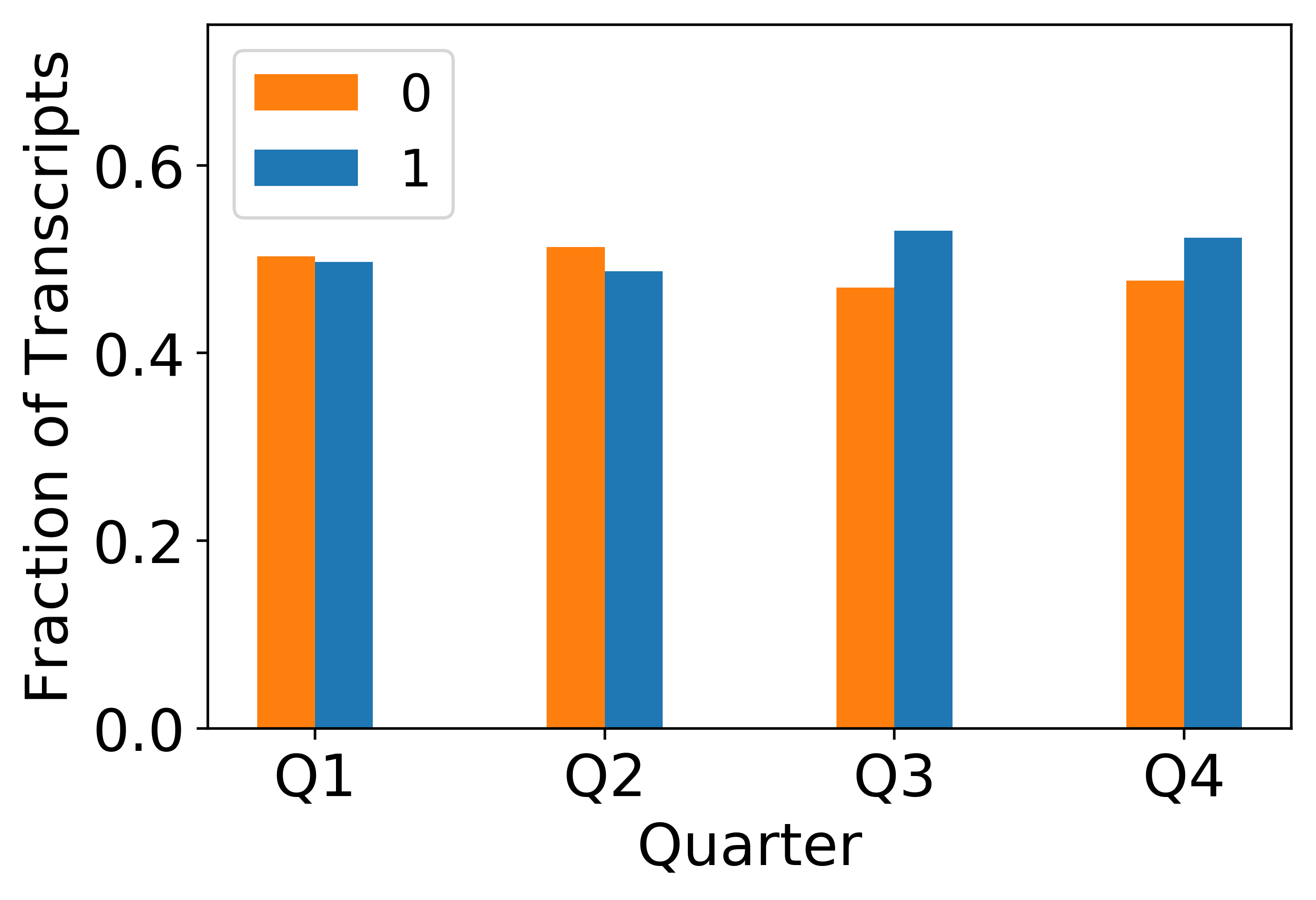}\label{fig:data_quar_daily}}
    \subfloat[ Year-wise for $y_s$]{\includegraphics[width=0.22\textwidth]{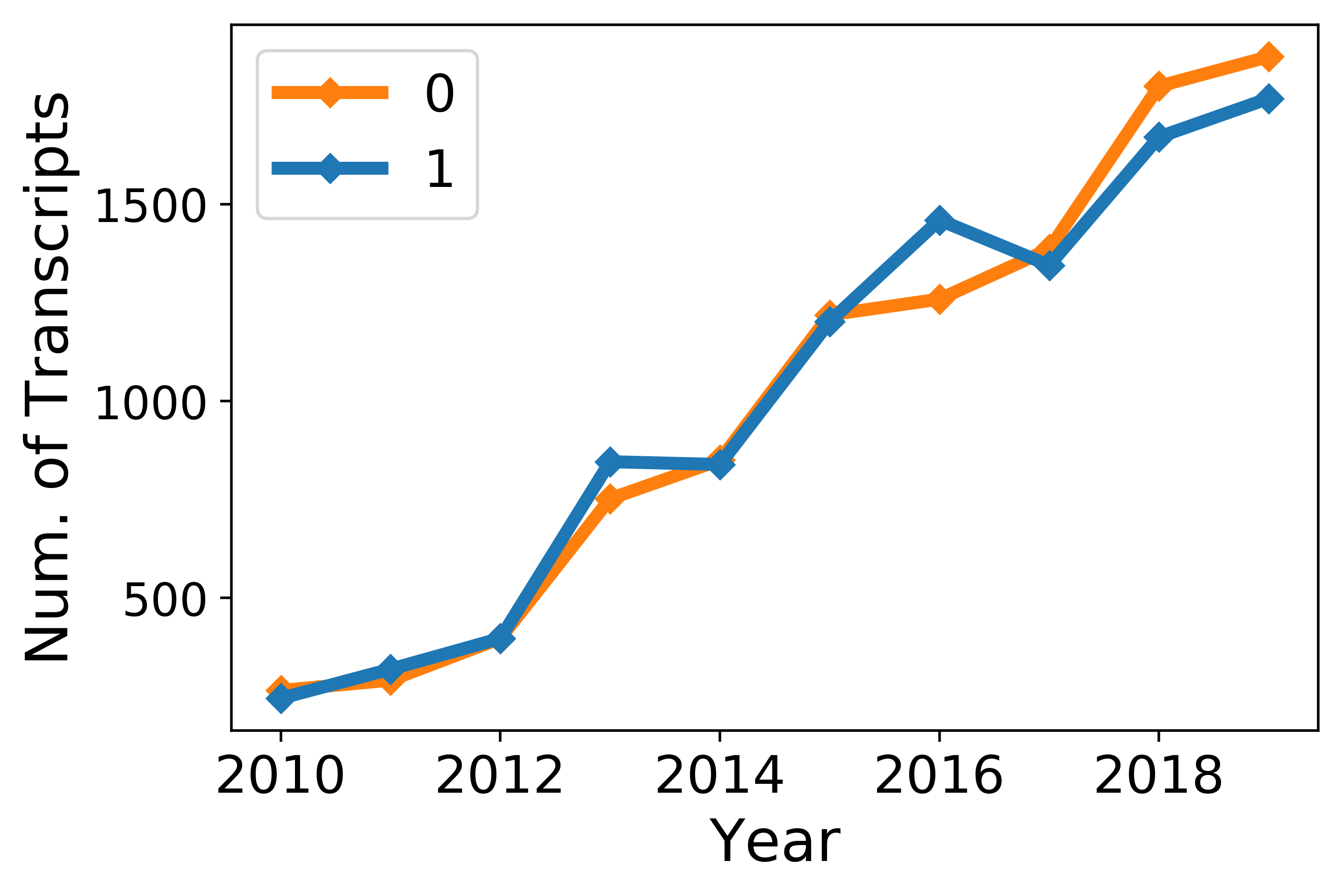}\label{fig:data_year_daily_shock}}
    \subfloat[Quarter-wise for $y_s$ ]{\includegraphics[width=0.22\textwidth]{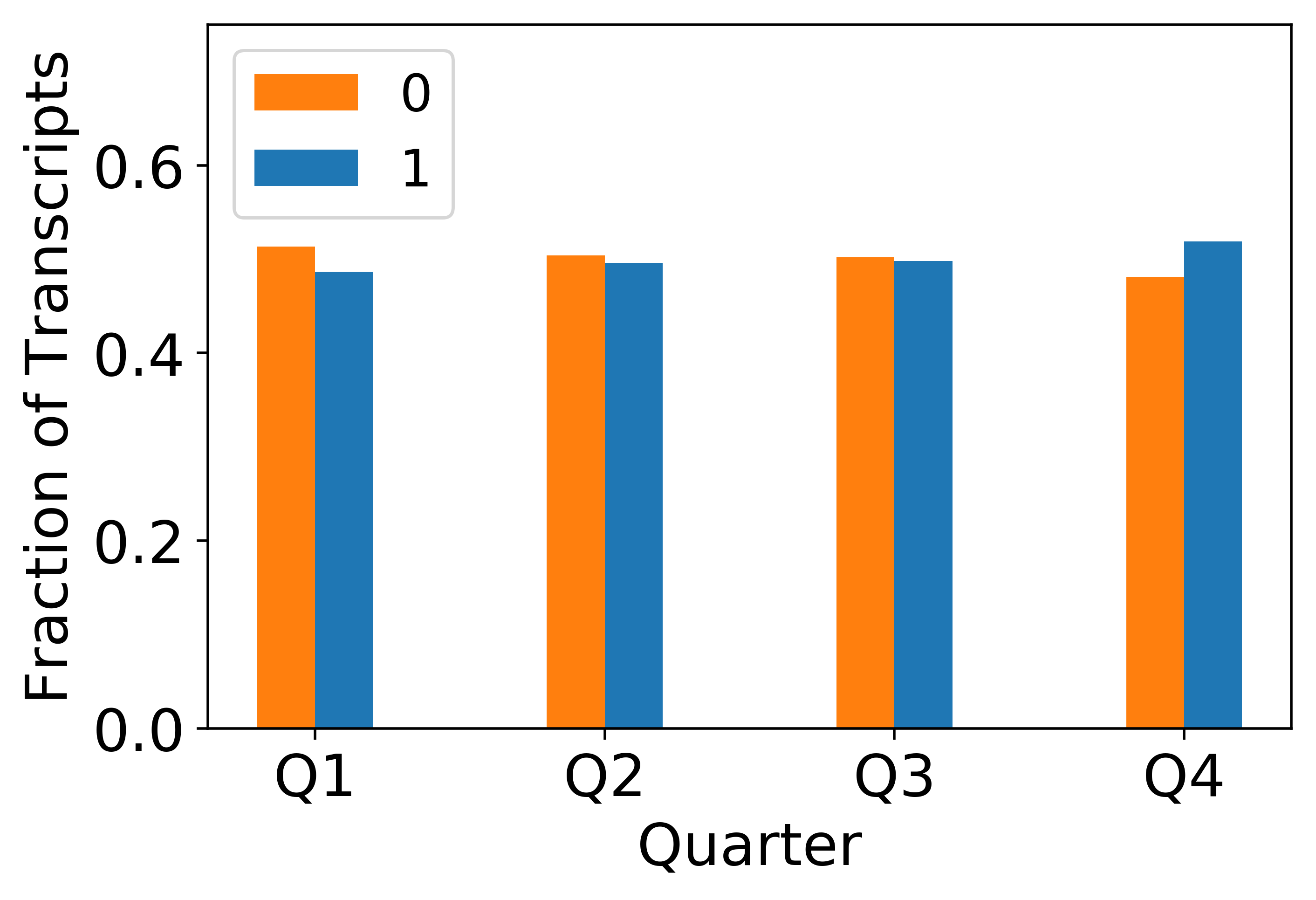}\label{fig:data_quar_daily_shock}}
    \vspace{-2mm}
    \caption{The (a) number of transcripts over the years and (b) fraction of transcripts in different quarters with both labels (orange for $0$ and blue for $1$) for the value based label ($y_v$). The (c) number of transcripts over the years and (d) fraction of transcripts in different quarters for the shock based label ($y_s$). } 
\label{fig:data_desc_year_quarter_daily_shock}
\end{figure*}

\begin{figure}[ht]
\vspace{-4mm}
    \centering
    \captionsetup[subfigure]{labelfont={normalsize,bf},textfont={normalsize,bf}}
    
    \subfloat[Year-wise for $y_{I,5}$]{\includegraphics[width=0.22\textwidth]{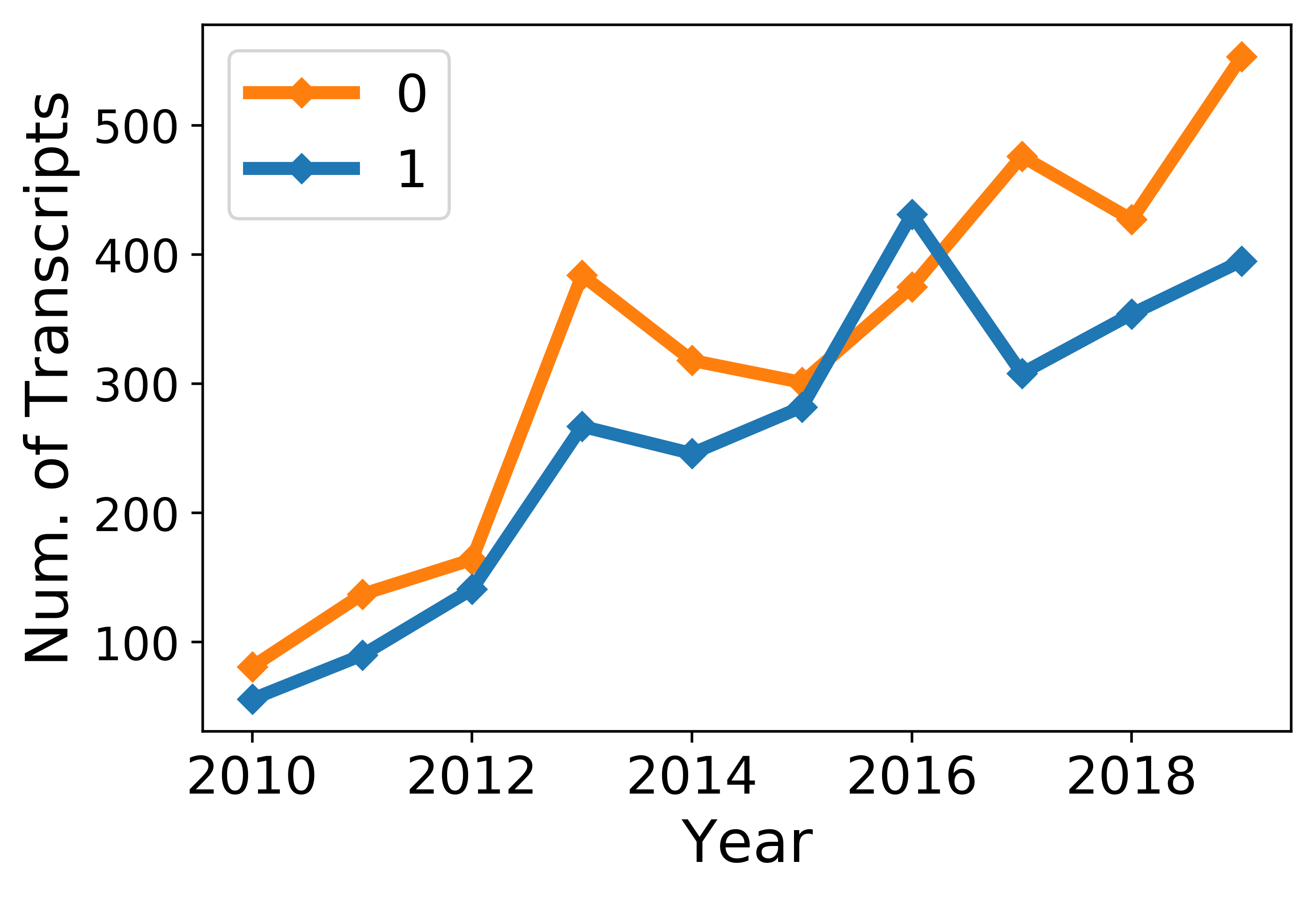}\label{fig:data_year_ind}}
    \subfloat[Quarter-wise for $y_{I,5}$]{\includegraphics[width=0.22\textwidth]{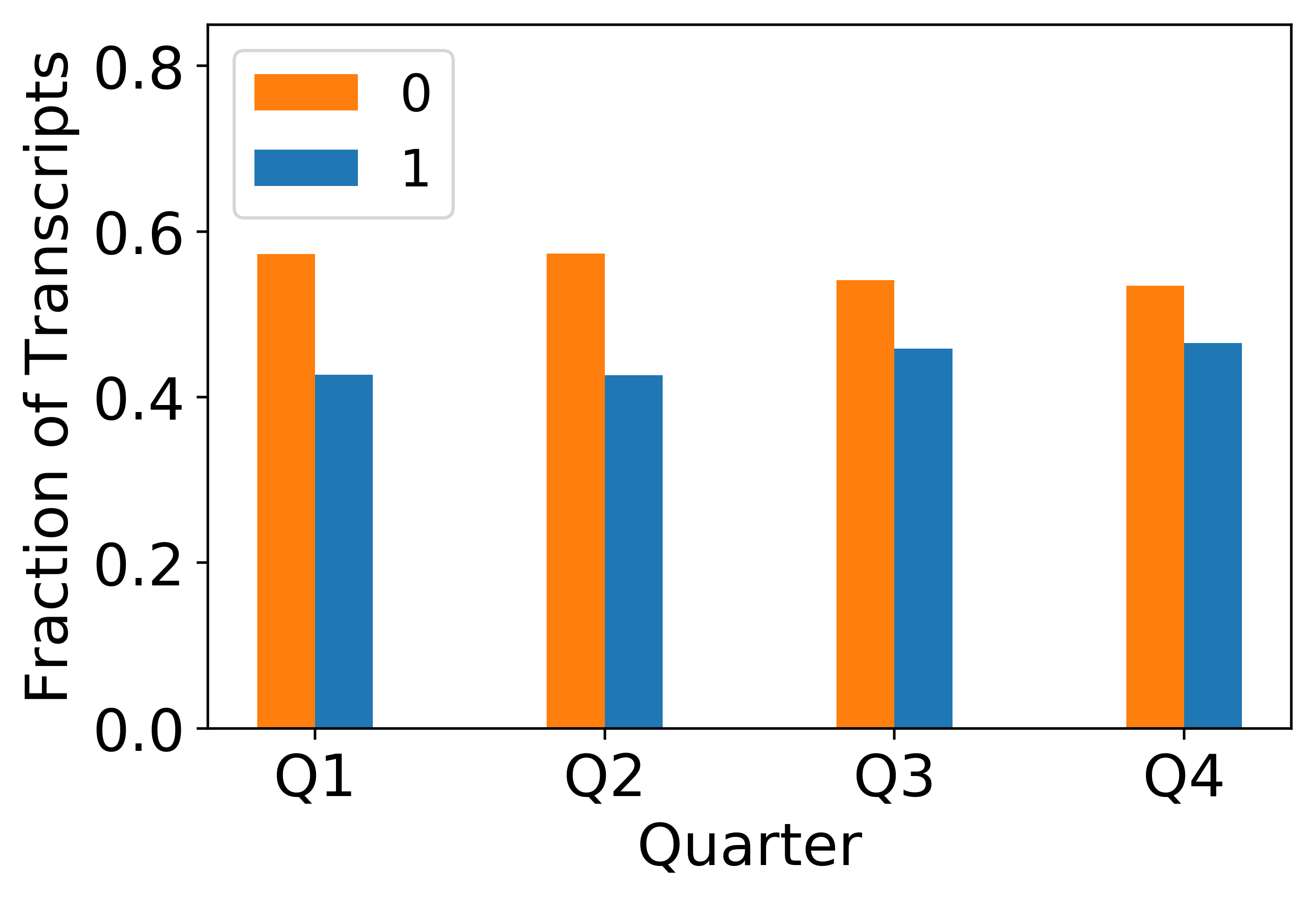}\label{fig:data_quar_ind}}
    \vspace{-1mm}
    \caption{ The (a) number of transcripts over the years and (b) fraction of transcripts in different quarters for the index based label ($y_{I,k}$ when $k=5$). } 
\label{fig:data_desc_year_quarter_index_5}
\end{figure}

\begin{figure}[ht]
\vspace{-8mm}
    \centering
    \captionsetup[subfigure]{labelfont={normalsize,bf},textfont={normalsize,bf}}
    
    \subfloat [$y_v$]{\includegraphics[width=0.16\textwidth]{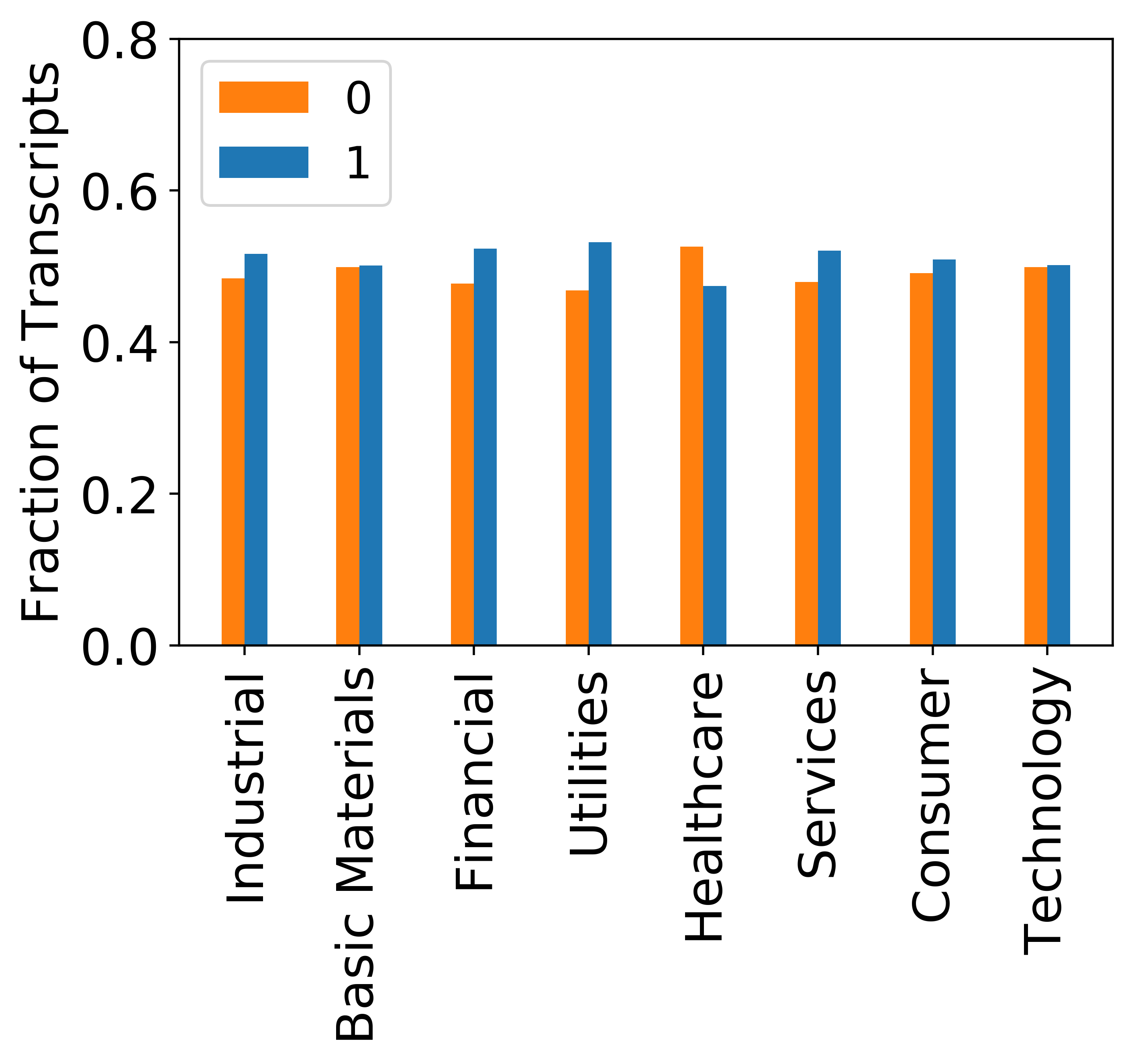} \label{fig:data_sector_value}}
    \subfloat[$y_s$]{\includegraphics[width=0.16\textwidth]{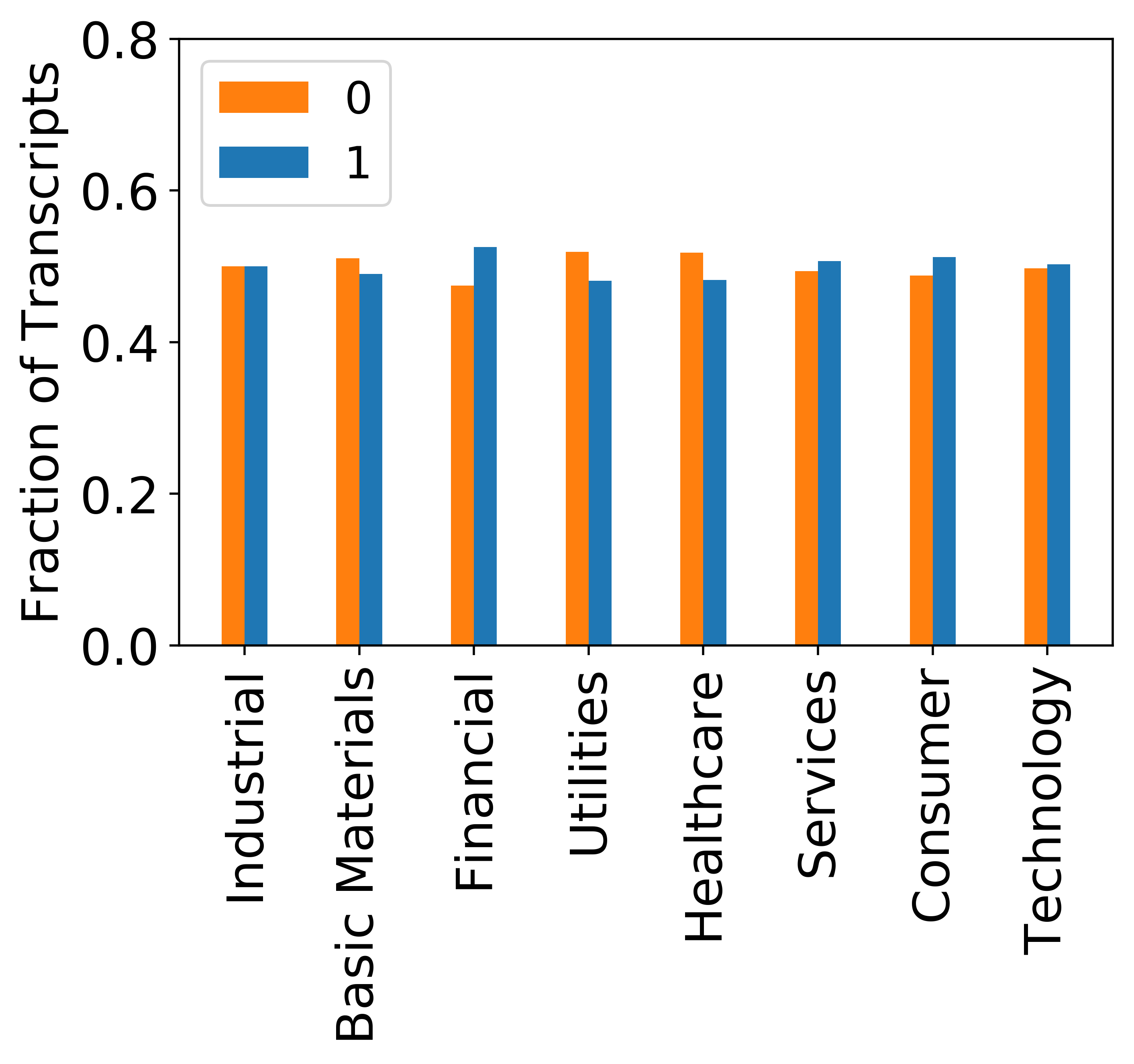} \label{fig:data_sector_value_shock}}
    \subfloat[$y_{I,5}$]{\includegraphics[width=0.16\textwidth]{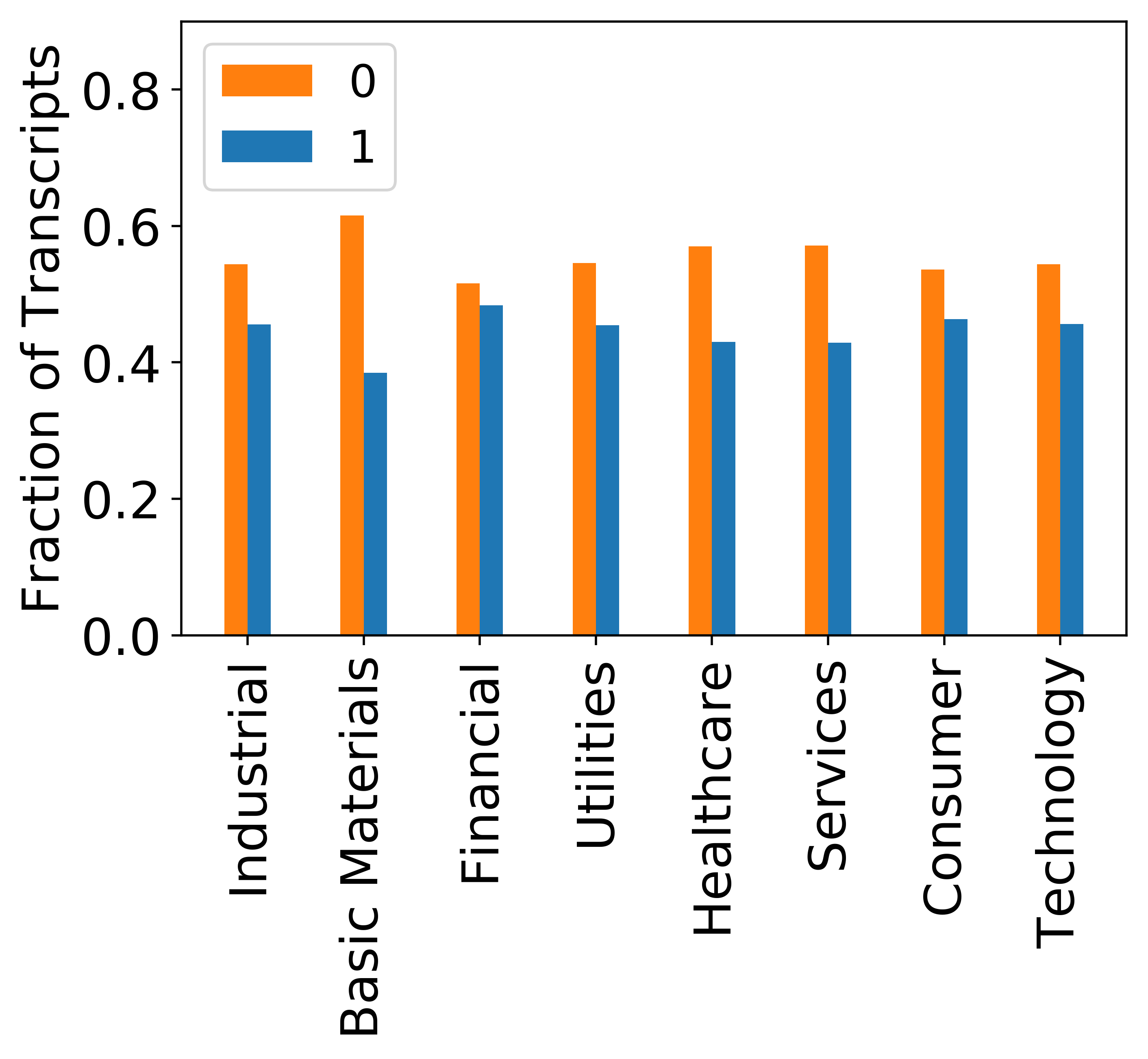}\label{fig:data_sector_index}}
    
\vspace{-2mm}
    \caption{The fraction of transcripts with both labels ($0$ and $1$) in different sectors when the labels are (a) value based ($y_v$) (b) shock based label ($y_s$) and (b) index based ($y_{I,5}$). } 
    \vspace{-1mm}
\label{fig:data_desc_area}
\end{figure}

\section{Data}
Our study includes three datasets--- earnings call transcripts, earnings per share (EPS) and sales estimates, and recommendations from analysts. We have collected transcripts that are available online\footnote{\url{https://seekingalpha.com/}}. The other datasets are from Zack investment\footnote{\url{https://www.zacks.com/}} dataset obtained via Wharton Research Data Services (WRDS)\footnote{\url{https://wrds-www.wharton.upenn.edu/pages/about/data-vendors/zacks/}}. We have collected full texts of 
152,483 transcripts of which 97,478 have complete data; the remaining transcripts were unusable due to missing data.
The earnings calls for these transcripts happen between January of 2010 to December of 2019. The dataset has all the relevant information such as the name of the company, ticker symbol, transcript release date, industrial sector, and quarter of the earnings. These transcripts are from 6,300 different companies that belong to one of nine predefined sectors as shown in Table \ref{tab:tab1}.

\subsection{Descriptive Analysis with Sentiments} We use the Linguistic Inquiry and Word Count (LIWC)\footnote{\url{https://www.kovcomp.co.uk/wordstat/LIWC.html}} dictionary to identify the sentiment traits in the transcripts. We include positive and negative emotions, anxiety, and sadness as affect traits among others. Figs. \ref{fig:sentiment_daily_years} and \ref{fig:sentiment_daily_years_shock} show the year-wise sentiments in the transcripts for $y_v$ (value label) and $y_s$ (shock label) respectively (the results with the index label ($Y_{I,5}$) are similar and omitted due to space). We make two interesting observations: (1) While the positive sentiment is higher for upward labels over the years; negative, sad and anxiety sentiments are higher for downward labels. (2) Over the years, the positive emotion has been increased and the negative and sad emotions have been decreased indicating that the earnings calls are being more enthusiastic over the years. We perform a detailed analysis of the role of sentiments in the stock price movement prediction in Sections \ref{sec_regression_analysis} and \ref{sec:senti_sem}.

\subsection{Descriptive Analysis on Label Distributions}
We show descriptive analysis of the data on the distribution of the labels. Figs. \ref{fig:data_desc_year_quarter_daily_shock} and \ref{fig:data_desc_year_quarter_index_5} represent the number and fraction of the transcripts in both classes ($0$ and $1$) over the years as well as in specific quarters as the earnings calls happen quarterly from a company.
Unsurprisingly, the number of transcripts increase  (Figs. \ref{fig:data_year_daily}, \ref{fig:data_year_daily_shock}, and \ref{fig:data_year_ind}) over the years for all label functions as the number of company increases. After aggregating the transcripts over different quarters, we find there is no significant difference in number of transcripts between two classes ($0$ and $1$) for the value and shock based labels (Figs. \ref{fig:data_sector_value} and \ref{fig:data_sector_value_shock}). However, this difference is more prominent in the index based label where class $0$ is larger (Fig. \ref{fig:data_sector_index}).  
\section{Pilot Regression Analysis}
In this section, we run several regression analyses to explore the association between text narratives and corporate stock market performance that are defined by $y_v$, $y_s$ and $y_{I,k}$ (see details in Section~\ref{sec:problem_definition}). 

\begin{table}[]
\footnotesize
\begin{tabular}{llllll}
Variables         & Model 1 & Model 2                                                    & Model 3                                                    & Model 4                                                     & Model 5                                                           \\
Positive Emotion  &         & \begin{tabular}[c]{@{}l@{}}0.26***\\ (0.012)\end{tabular}  & \begin{tabular}[c]{@{}l@{}}0.26***\\ (0.013)\end{tabular}  & \begin{tabular}[c]{@{}l@{}}0.26***\\ (0.013)\end{tabular}   & \begin{tabular}[c]{@{}l@{}}0.29***\\ (0.018)\end{tabular}         \\
Negative Emotion  &         & \begin{tabular}[c]{@{}l@{}}-0.64***\\ (0.033)\end{tabular} & \begin{tabular}[c]{@{}l@{}}-0.33***\\ (0.057)\end{tabular} & \begin{tabular}[c]{@{}l@{}}-0.34***\\ (0.057)\end{tabular}  & \begin{tabular}[c]{@{}l@{}}-0.32***\\ (0.077)\end{tabular}        \\
Anxiety Score     &         &                                                            & \begin{tabular}[c]{@{}l@{}}-0.26*\\ (0.11)\end{tabular}    & \begin{tabular}[c]{@{}l@{}}-0.24*\\ (0.11)\end{tabular}     & \begin{tabular}[c]{@{}l@{}}-0.19\\ (0.16)\end{tabular}            \\
Anger Score       &         &                                                            & \begin{tabular}[c]{@{}l@{}}0.17\\ (0.12)\end{tabular}      & \begin{tabular}[c]{@{}l@{}}0.19\\ (0.12)\end{tabular}       & \begin{tabular}[c]{@{}l@{}}0.19\\ (0.16)\end{tabular}             \\
Sad Score         &         &                                                            & \begin{tabular}[c]{@{}l@{}}-0.73***\\ (0.091)\end{tabular} & \begin{tabular}[c]{@{}l@{}}-0.75***\\ (0.092)\end{tabular}  & \begin{tabular}[c]{@{}l@{}}-1.22***\\ (0.13)\end{tabular}         \\
Certainty Score   &         &                                                            & \begin{tabular}[c]{@{}l@{}}-0.0097\\ (0.032)\end{tabular}  & \begin{tabular}[c]{@{}l@{}}0.029\\ (0.036)\end{tabular}     & \begin{tabular}[c]{@{}l@{}}0.061\\ (0.050)\end{tabular}           \\
Cognitive Score   &         &                                                            &                                                            & \begin{tabular}[c]{@{}l@{}}-0.019**\\ (0.0075)\end{tabular} & \begin{tabular}[c]{@{}l@{}}-0.034**\\ (0.011)\end{tabular}        \\
Insight Score     &         &                                                            &                                                            & \begin{tabular}[c]{@{}l@{}}-0.020\\ (0.021)\end{tabular}    & \begin{tabular}[c]{@{}l@{}}-0.038\\ (0.029)\end{tabular}          \\
Causation Score   &         &                                                            &                                                            & \begin{tabular}[c]{@{}l@{}}-0.024\\ (0.021)\end{tabular}    & \begin{tabular}[c]{@{}l@{}}0.021\\ (0.029)\end{tabular}           \\
Discrepancy Score &         &                                                            &                                                            & \begin{tabular}[c]{@{}l@{}}0.062*\\ (0.029)\end{tabular}    & \begin{tabular}[c]{@{}l@{}}0.027\\ (0.042)\end{tabular}           \\
Actual Sales      &         &                                                            &                                                            &                                                             & \begin{tabular}[c]{@{}l@{}}8.33e-05***\\ (2.36e-05)\end{tabular}  \\
Estimated Sales   &         &                                                            &                                                            &                                                             & \begin{tabular}[c]{@{}l@{}}-8.71e-05***\\ (2.42e-05)\end{tabular} \\
Actual EPS        &         &                                                            &                                                            &                                                             & \begin{tabular}[c]{@{}l@{}}0.0056\\ (0.0041)\end{tabular}         \\
Estimated EPS     &         &                                                            &                                                            &                                                             & \begin{tabular}[c]{@{}l@{}}-0.0059\\ (0.0043)\end{tabular}        \\
Sector            & YES     & YES                                                        & YES                                                        & YES                                                         & YES                                                               \\
Year              & YES     & YES                                                        & YES                                                        & YES                                                         & YES                                                               \\
Observations      & 93,682  & 93,165                                                     & 93,165                                                     & 93,165                                                      & 51,306           \\                                                
BIC               & 129,921 & 128,040                                                    & 127,968                                                   & 127,994                                                   & 70,465                                                      
\end{tabular}
\caption{Logistic Regression of Value Based Label Function ($y_v$) on sentiment features. Standard errors are in parentheses.  Significance levels marked with stars: *** p<0.001, ** p<0.01, * p<0.05. YES denotes inclusion of the variable. Bayesian Information Criterion (BIC) \cite{RePEc:cup:cbooks:9780521852258} is used for model selection and lower BIC is preferred.}
\vspace{-6mm}
\label{tab:daily_movement_prediction}
\end{table}

\subsection{Effect of Sentiments}
\label{sec:regress_senti}
Our objective is to validate whether information in the text of the earnings call transcripts have predictive power after controlling for several frequently used performance indicators of corporate’s performance, such as earnings per share (EPS) and sales (see details in Section \ref{sec:back_sales}).

Here, we use $y_v$, $y_s$ and $y_{I,k}$ as our dependent variables in the regression (logistic). Our key independent variables are quantitative measures (e.g., sentiments) extracted from the transcripts. We quantify sentiments using LIWC dictionary. There are two categories of them. First, we have major metrics, which quantify the levels of positive and negative emotion along with more detailed emotion information, such as anxiety, anger, and sadness. Second, we also measure personality features in the texts, such as how certain the language is, how many insight words are used, and how many cognitive processes words are used in the speech among others.

In Table~\ref{tab:daily_movement_prediction}, we present the estimates (coefficients, standard errors, significance) of the variables in five regression models where $y_v$ is the dependent variable along with \textit{year} and \textit{sector} as fixed effects. We have several interesting observations. First, we find that positive emotion and negative emotion are significantly predictive of stock market performance in terms of $y_v$ (model 2 in Table~\ref{tab:daily_movement_prediction}). Second, both of them remain significant after controlling for corporate’s actual sale, estimated sale, actual EPS, and estimated EPS. This implies that the sentiment information extracted from texts can provide useful signal for stock price movements other than frequently used indicators (e.g., sales, EPS). Thirdly, the signs of positive sentiment and negative sentiment are in the opposite directions. This means that earnings call transcripts with positive emotion are predicting stock price increase. On the other hand, the amount of negative emotion in the transcripts correlates with stock price decrease. Similar results are found for sadness score and cognitive process words. In Appendix (Table~\ref{tab:shock_label_prediction}), we present the same analyses for the shock based label $y_s$ as dependent variable. All results are consistent with our above observations.

To summarize, this pilot analysis using metrics from texts indicate a connection between transcript narratives and stock market performance. This leads us to design more rigorous NLP methods based on semantics (Sec. \ref{sec:method}).

\subsection{Sentiment of Transcripts and Analysts' Recommendations }
\label{sec_regression_analysis}
We explore the effect of recommendations from the analysts in this section for stock price movement prediction. Section \ref{sec_regression_results} shows that while analysts' recommendations prior to earnings calls are not significant, the sentiment traits of the transcripts are important. 

\subsubsection{Settings}
\label{sec_sentiment_def}



Analysts play a role in affecting the perceptions of the investors about the future prospects of a business \cite{barber2001can,francis1997relative,asquith2005information} and thus their recommendations can affect the stock price. We aim to explore the association between the analysts' recommendations and our proposed value based ($y_v$) label (the results on index ($y_{I,k}$) based label are in Sec. \ref{sec:extra_analyst} of the Appendix). The recommendations are captured in a categorical variable with five possible values: strong buy, moderate buy, hold, moderate sell, and strong sell. The number of recommendations ranges from 10 to 500 for each company.


To measure the extent of the associations between the recommendations and the earnings calls, the dates of occurrences are important. The analysts' recommendations are not always available on the dates of the earnings calls. Thus, we create variables correspond to the ``majority recommendation'' on a company within time $t$ prior to the date of the earnings call. For instance, if $t$ is 1 month, a company's earnings calls date is 2015-09-20, 10 recommendations are made since 2015-08-20, and six of them (or the majority) are marked as ``hold''; the value of the recommendation variable corresponding to this earnings call will be ``hold''. In our analysis we set $t$ as one month and denote this Monthly Analysts' Recommendation (MAR) variable as $MAR_{1m}$. 
 Additionally, we create another variable to measure the ``local after-affect''. In particular, we aggregate the recommendations from the analysts within 5 days after the earnings calls date and assign the values with majority recommendations as before. We denote it as $MAR_{5d}$.

\begin{table}[t]
\footnotesize
\begin{tabular}{|l|c c c c c|}
\hline
Variables & Model   1 & Model   2 & Model   3 & Model   4 & Model   5 \\ \hline
positive  & 0.27***   & 0.27***   & 0.26***   & 0.27***   & 0.34***   \\ 
          & (0.03)    & (0.03)    & (0.03)    & (0.03)    & (0.04)    \\ \hline
negative  & -0.43***  & -0.43***  & -0.42***  & -0.42**   & -0.63***  \\ 
          & (0.12)    & (0.12)    & (0.13)    & (0.13)    & (0.19)    \\ \hline
anxiety   & 0.35      & 0.35      & 0.28      & 0.28      & 0.07      \\ 
          & (0.23)    & (0.23)    & (0.24)    & (0.24)    & (0.35)    \\ \hline
anger     & 0.56*     & 0.54*     & 0.57*     & 0.56*     & 0.62      \\ 
          & (0.24)    & (0.24)    & (0.25)    & (0.25)    & (0.36)    \\ \hline
sad       & -0.66***  & -0.58**   & -0.63**   & -0.63**   & -0.57     \\ 
          & (0.19)    & (0.19)    & (0.20)    & (0.20)    & (0.29)    \\ \hline
certain   & -0.01     & -0.02     & -0.01     & -0.00     & 0.12      \\ 
          & (0.07)    & (0.07)    & (0.07)    & (0.07)    & (0.11)    \\ \hline
$MAR_{1m}$  &           & YES       & YES       & YES       & YES       \\ \hline
Sector  &           &           & YES       & YES       & YES       \\ \hline
Year      &           &           &           & YES       & YES       \\ \hline
$MAR_{5d}$  &           &           &           &           & YES       \\ \hline
Observations        & 18,289    & 18,289    & 18,289    & 18,289    & 8,994     \\ 
BIC       & 25185     & 25201     & 25271     & 25346     & 12474     \\ \hline
\end{tabular}
\caption{Regression Results: Value Based Label Function ($y_v$) on Analysts' Recommendations within 1 month ($MAR_{1m}$) prior and 5 days ($MAR_{5d}$) after the earnings call date and sentiment of the transcripts. Standard errors are shown in parentheses for the coefficients. Significance levels marked with stars: *** p<0.001, ** p<0.01, * p<0.05. YES denotes the inclusion of the variable. }
\label{reg_mar_3_1_value}
\vspace{-5mm}
\end{table}



\subsubsection{Effect of Sentiments and Recommendations}
\label{sec_regression_results}

We attempt to understand the associations among analysts recommendations, sentiment of the earnings call transcripts and our defined labels (classes). We use $p$-value \cite{wasserstein2016asa} to evaluate the  significance of a variable. Bayesian Information Criterion (BIC) \cite{RePEc:cup:cbooks:9780521852258} is used for the performance measure in model selection and the model with the lowest BIC is preferred.

The key independent variables are analysts' recommendations and sentiment of earnings call transcripts. For analysts' recommendations, the variables $MAR_{1m}$ and $MAR_{5d}$ describe analysts' ratings one month prior to and five days after the day of earnings calls respectively. 
There are six variables related to sentiments of transcripts: $positive$, $negative$, $anxiety$, $anger$, $sad$, and $certain$.

\textbf{Results: }Table~\ref{reg_mar_3_1_value} shows the regression results for $y_{v}$. Model 1 only takes six sentiment variables as independent ones.  In Model 2, we add analysts recommendations ($MAR_{1m}$) as another independent variable. Similarly, we add fixed effect of the areas of the companies, fixed effect of the year of the earnings calls, and analysts' recommendations after Earnings call ($MAR_{5d}$) in Model 3, Model 4 and Model 5 respectively. The values in the same row indicate the coefficients of the variables with significance levels ($p$-values).

The results on the sentiments are consistent with the ones in Sec. \ref{sec:regress_senti}. Moreover, adding prior recommendations analysts, industry information and year do not improve the model's fitness as the BIC values in Model 2, 3 and 4 are larger (smaller means better) than that of Model 1. However, Model 5 has a much lower BIC value. The reasons could be the followings: the sample size ($N$) is different; and Model 5 includes posterior analysts' recommendations $MAR_{5d}$, which intuitively should have better explanation power of the dependent variable $y_{v}$.

\textbf{Implications:}
There are two major implications of these results. 
\begin{itemize}
    \item The ratings from the analysts, that are made prior to the earnings call do not have significant impact on the stock price movement (SPM) prediction after the earnings call. These  ratings are usually categorical variables and thus, these might not translate directly to the actions that investors might take and affect SPM. However, sentiments can be easily visible or interpretable to the investors through special adjective or words. 
    
    \item As analysts are domain experts, they usually have a global viewpoint on the growth of the company. However, our defined binary classification tasks capture a localized trend after the occurrence of the earnings call. Therefore, global viewpoints that are made prior to it, might not be useful to capture such local effects.
\end{itemize}

\section{Earnings calls and stock price movement prediction}
\label{sec:method}
Our goal is to learn prediction functions where the features are from the earnings call transcripts with value based ($y_v$), shock based ($y_s$) and index based $y_{I,k}$ labels. 
We begin with descriptions of our approaches to solve these classification problems. We show the main results in Sec. \ref{sec:acc_results} and a comparison against a baseline consisting of ``hard" data such as earnings per share (EPS) and sales in Sec. \ref{sec:vs_SE}.

\subsection{Our Method}
\label{sec:our_method}
 To test the predictive power of the content in the earnings call, we use their semantics to solve the aforementioned problems. Our main method \textsc{StockGNN} is a combination of a graph neural network (GNN) architecture \cite{kipf2016semi} and a semantic feature generator by the well-known Doc2Vec method \cite{le2014distributed}.
 
\begin{figure}[t]
    \centering
    
    {\includegraphics[width=0.5\textwidth]{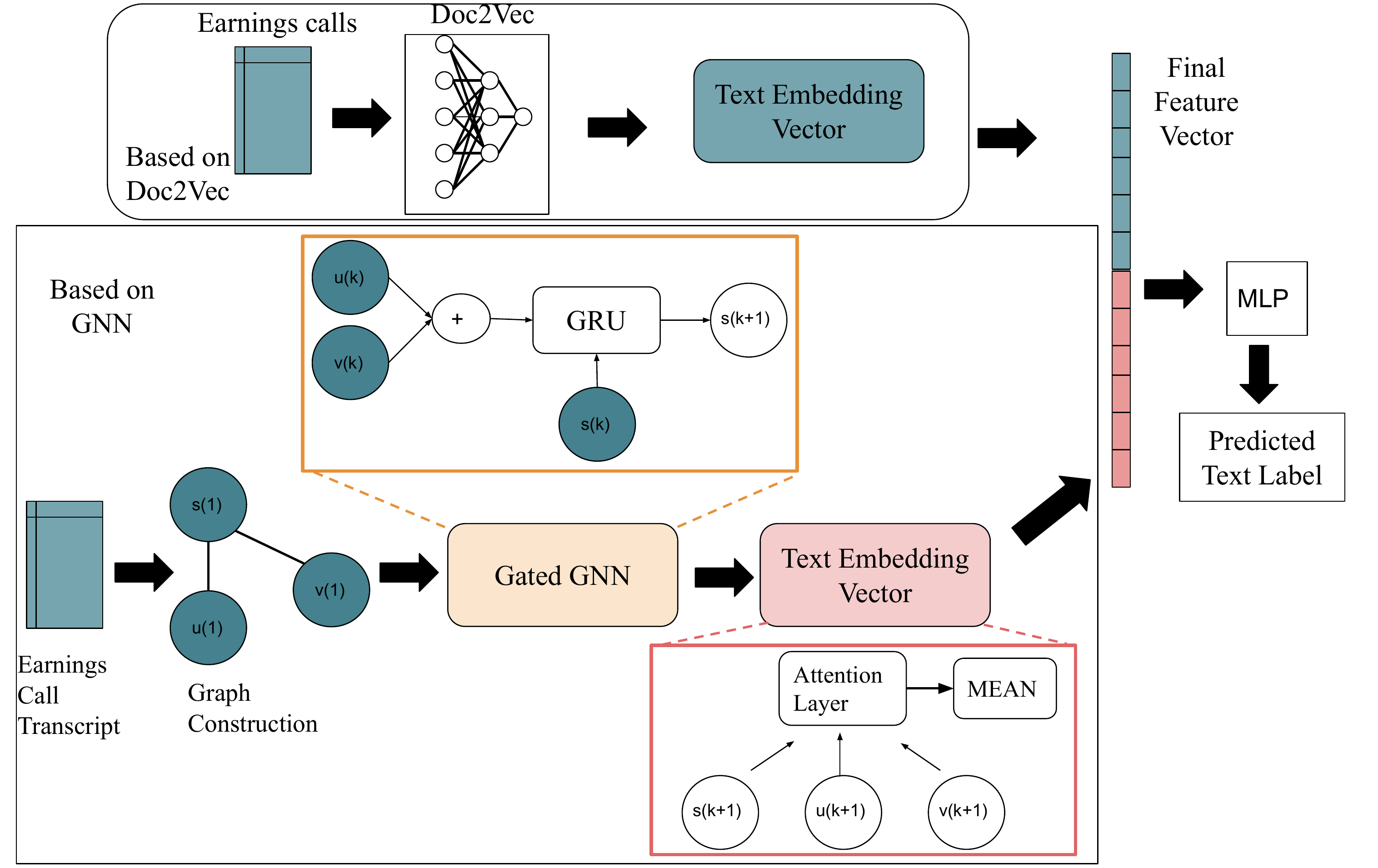}}
   
\vspace{-3mm}
    \caption{The architecture of  \textsc{StockGNN}: It combines traditional context based Doc2Vec embeddings with GNN based embeddings that capture text level word interactions.  } 
\label{fig:architecture}
\vspace{-1mm}
\end{figure}

 \textbf{The \textsc{StockGNN} Method:} 
 GNN based methods have been popular tools in several natural language processing and related tasks such as event detection \cite{deng2019learning,huang2019text,zhang2020every,kosan2021event}. Inspired by the method in \cite{zhang2020every}, our architecture (Figure \ref{fig:architecture}) is based on a Gated GNN \cite{li2015gated} and have four following components:\\ \textbf{1) Graph generation:} We first build a graph $G=(V,E)$ from each document. Each unique word in that document becomes a node and the words appear in its neighbourhood (or context) become its neighbors. The neighbourhood size becomes a hyper-parameter.  \\ \textbf{2) Gated GNN:} After building the graph, a Gated GNN is applied to learn the node (word) embeddings. We start with initial (given or constructed features) embeddings ($z\in \mathbb{R}^{|V|\times d}$, a vector of size $d$) of the nodes or words. The gated GNN produces an embedding of node $x$ based on aggregation of its own features and its neighbours' features. To achieve higher order ($k$-hop neighbours) interactions, this process is repeated $k$ times. More specifically, the Gated GNN \cite{li2015gated} interactions are captured by the following equations:
 \begin{equation*}
 \begin{aligned}
x^k = \sigma(\textbf{W}_xa^k +\textbf{U}_xz^{k-1}+b_x)\text{ where }a^k = \textbf{A}z^{k-1}\textbf{W}_a\\
r^k = \sigma(\textbf{W}_ra^k +\textbf{U}_rz^{k-1}+b_r)\\
\hat{z}^k =tanh(\textbf{W}_za^k+ \textbf{U}_z(r^k \otimes z^{k-1})+b_z)\\
z^k = \hat{z}^k+ z^{k-1} \otimes (1-x^k)
\end{aligned}
\end{equation*}
Here, $\otimes$ is element-wise multiplication, $\sigma$ is the sigmoid function, $z^k$ is the final embedding after $k$-hop interactions, \textbf{A} is the adjacency matrix of the graph, $\textbf{W}, \textbf{U}$ and $b$ are trainable weights (parameters) and biases. Furthermore, $x$ and $r$ are update and reset gate respectively and they are used to determine the degree of information from the neighbours and the current node into consideration. As the words in the same document form a graph,  Gated GNN produces embeddings based on both sentence (neighbours are created from context window) and document level interactions.\\
 \textbf{3) Embedding vector:} The classification label is associated with the document, i.e., the corresponding graph. So the node embeddings learned via the Gated GNN are further aggregated by a  function and generate final embeddings ($z_G$) for the entire graph or document. The aggregation function is as follows:
 \begin{equation*}
 \begin{aligned}
z_G = \frac{1}{|V|} \sum_{v\in V}z_v\text{ where }z_v = \sigma(MLP(z^k_v)) \otimes tanh(MLP(z^k_v))
\end{aligned}
\end{equation*}
where $z_v$ is embedding for the node $v$ and MLP is Multilayer Perceptron. Afterwards the embeddings are fed into a softmax layer and trained by a cross-entropy function. \\
\textbf{4) Combining embeddings via Doc2Vec and final classification:} We further generate unsupervised embeddings via the well-known Doc2Vec method \cite{le2014distributed} and concatenate with the (supervised) embeddings from the previous step. Thus, we exploit having both unsupervised embeddings based on content similarity of transcripts from Doc2Vec and sophisticated supervised ones from Gated GNN that produces embeddings based on both sentence and document level interactions. These final embeddings are then fed into a MLP to generate the final classification. 

The main advantages of \textsc{StockGNN} are threefold: (i) easy to implement, (ii) captures both sentence and document level interactions, (iii) can be generalized with other GNN architectures. In experiments, we show that \textsc{StockGNN} also predicts relatively accurate daily and weekly stock price movements with all labels. 
 \label{sec:methods}

\begin{table*}[ht]
    \centering
    \small
\vspace{-3mm}
\begin{tabular}{| c||c c c c c|| c c c c c||c c c c c||}
\hline
\textbf{Measures}&\multicolumn{5}{|c||}{\bf Accuracy}&\multicolumn{5}{|c||}{\bf Avg. Precision}& \multicolumn{5}{c||}{\bf Avg. Recall}\\
\hline
\textbf{Methods} & Fin & Health & Mat & Service & Tech & Fin & Health & Mat & Service & Tech & Fin & Health & Mat & Service & Tech \\
\hline
\textsc{DEsvm} & .624 & .549 & .74 & .647 & .527 & .633& .533 & .661 & .58& .529 & .632 & .526& .605 & \textbf{.592} & .525\\
\textsc{DElogreg} & .595 & .524 & .632 & .630 & .534 & .609& .516 & .584 & .57& .542 & .606 & .516& .610& .56 & .545\\
\textsc{DEmlp} & \textbf{.666} & .540 & .724 & .595 & .565 & \textbf{.664}& .524 & .615 & .564& .538 & \textbf{.665} & .520& .575 & .568 & .539\\
\textsc{StockGNN} & .582 & \textbf{.60} & \textbf{.761} & \textbf{.65} & \textbf{.583} & .631& \textbf{.60} & \textbf{.69} & \textbf{.581}& \textbf{.554} & .61 & \textbf{.578}& \textbf{.614} & .55 & \textbf{.553}\\
\hline

\end{tabular}
\caption{\textbf{Index Based Label ($y_{I,5}$) Results:} The accuracy, average precision and recall produced by different methods in five major sectors or areas. The maximum standard deviations in the results for \textsc{DEsvm}, \textsc{DElogreg}, \textsc{DEmlp}, and \textsc{StockGNN} are $.002, .002, .008$ and $.003$ respectively. The best performances are shown in bold. Our main method \textsc{StockGNN} outperforms the baselines in most of the cases. A base rate (label all data as the label of the larger class) will produce an average recall of .5. }
\label{tab:results_strong_week}

 \end{table*}

 
\begin{table*}[ht]
    \centering
    \small
\vspace{-5mm}
\begin{tabular}{| c||c c c c c|| c c c c c||c c c c c||}
\hline
\textbf{Measures}&\multicolumn{5}{|c||}{\bf Accuracy}&\multicolumn{5}{|c||}{\bf Avg. Precision}& \multicolumn{5}{c||}{\bf Avg. Recall}\\
\hline
\textbf{Methods} & Fin & Health & Mat & Service & Tech & Fin & Health & Mat & Service & Tech & Fin & Health & Mat & Service & Tech \\
\hline
\textsc{DEsvm} & .563 & .603 & .52 & \textbf{.57} & .57 & \textbf{.62} & .61 & .52 & .57 & .58 & .563 & \textbf{.61}& .52 & \textbf{.57} & .58\\
\textsc{DElogreg} & .551 & .59 & .52 & .56 & .58 & .555& .59 & .516 & .56& .58 & .555 & .59& .51& .56 & .58\\
\textsc{DEmlp} & .592 & .57 & .53 & .54 & .60 & .587 & .57 & .52 & .54& .60 & .585 & .57& .52 & .54 & .60\\
\textsc{StockGNN} & \textbf{.60} & \textbf{.61} & \textbf{.62} & .568 & \textbf{.61} & .60 & \textbf{.61} & \textbf{.62} & \textbf{.57} & \textbf{.61} & \textbf{.602} & .604 & \textbf{.62} & \textbf{.57} & \textbf{.613}\\
\hline

\end{tabular}
\caption{\textbf{Shock Based Label ($y_{s}$) Results:} The accuracy, average precision and recall produced by different methods. The maximum standard deviations in the results for \textsc{DEsvm}, \textsc{DElogreg}, \textsc{DEmlp}, and \textsc{StockGNN} are $.003, .001, .007$ and $.004$ respectively. The best performances are shown in bold. Our main method \textsc{StockGNN} outperforms the baselines in most of the cases. }
\label{tab:results_daily_shock}

 \end{table*}

\begin{table*}[ht]
    \centering
    \small
\vspace{-5mm}
\begin{tabular}{| c||c c c c c|| c c c c c||c c c c c||}
\hline
\textbf{Measures}&\multicolumn{5}{|c||}{\bf Accuracy}&\multicolumn{5}{|c||}{\bf Avg. Precision}& \multicolumn{5}{c||}{\bf Avg. Recall}\\
\hline
\textbf{Methods}& Fin & Health & Mat & Service & Tech & Fin & Health & Mat & Service & Tech & Fin & Health & Mat & Service & Tech \\
\hline
\textsc{DEsvm} & .544 & .582 & .554 & .567 & .597 & .54 & .577 & .555 & .568& .597 & .54 & .579& .555 & .567 & .598\\
\textsc{DElogreg} & .55 & .584 & .556 & .565 & \textbf{.598} & .55 & .58 & .56 & .57 & .59 & .54 & .58 & \textbf{.56} & .57 & \textbf{.60}\\
\textsc{DEmlp} & .547 & .552 & .55 & \textbf{.574} & .549 & .54 & .56 & .55 & \textbf{.58} & .55 & .541 & .55 & .55 & \textbf{.574} & .55\\
\textsc{StockGNN} & \textbf{.638} & \textbf{.606} & \textbf{.563} & .56 & .55 & \textbf{.62} & \textbf{.609} & \textbf{.562} & .56& \textbf{.603} & \textbf{.544} & \textbf{.608}& \textbf{.56} & .545 & .562\\
\hline

\end{tabular}
\caption{\textbf{Value Based Label ($y_v$) Results:} The accuracy, average precision and recall by different methods in five major sectors. The maximum variances in the results for \textsc{DEsvm}, \textsc{DElogreg}, \textsc{DEmlp},  and \textsc{StockGNN} are $.001, .002, .008$ and $.005 $ respectively. The best performances are shown in bold. Our main method \textsc{StockGNN} outperforms the baselines in most of the cases.} 
\label{tab:results_value_label}
\vspace{-2mm}
 \end{table*}

 \begin{table}[ht]
\centering
\small
\begin{tabular}{| c||c c || c c || c c||}
\hline
\textbf{Labels}&\multicolumn{2}{|c||}{\bf VBL ($y_v$)}&\multicolumn{2}{|c||}{\bf SBL ($y_s$)}&\multicolumn{2}{|c||}{\bf IBL ($y_{I,5}$)}\\
\hline
\textbf{Areas} & Train & Test & Train & Test & Train & Test \\
\hline
\textsc{Fin} & 15000 & 2100 & 1052 & 174 & 909 & 141 \\
\textsc{Health} & 9839 & 1734 & 3152 & 673 & 657 & 122 \\
\textsc{Mat} & 9295 & 1300 &1425 & 252 & 603 & 87 \\
\textsc{Service} & 14990 & 2181 & 3511 & 615& 946 & 146 \\
\textsc{Tech} & 14182 & 2397 & 5000 & 968 & 896 & 161 \\
\hline

\end{tabular}
\caption{The number of transcripts in specific sectors (areas) with the value ($y_v$), shock ($y_s$), and index ($y_{I,5}$) based  labels. }
\label{tab:train_test_both_label}
\vspace{-5mm}
 \end{table}

\textbf{Other methods: }Our aim is to show the predictive power of the semantic features constructed from the transcripts as well as to measure the accuracy of the proposed \textsc{StockGNN} method compared to baselines. To achieve that we propose several baselines which have two phases. First, we generate unsupervised low-dimensional embeddings via the well-known Doc2Vec method \cite{le2014distributed}. Intuitively, similar documents will be embedded closer in that low-dimensional space. Second, these embeddings are further fed into a classifier for the final classification or label prediction. We have used three different classifiers: Support Vector Machine (SVM), Logistic Regression (logreg), and a Multilayer Perceptron (MLP). Other classifiers (such as Naive Bayes, k-NN) produce inferior results. We call the entire pipeline of using Doc2Vec and a classifier as \textsc{DEsvm}, \textsc{DElogreg}, and \textsc{DEmlp} when the classifiers are SVM, logreg and MLP respectively. \textsc{DEmlp} is a modified version of the method proposed in \cite{ma2020earnings}. Unlike using prior embeddings of words and aggregating them, we use Doc2Vec \cite{le2014distributed} to generate the embeddings for the document. Furthermore, we construct features only from transcripts as our aim is to classify with features from transcripts, whereas the method in \cite{ma2020earnings} use the company embeddings as additional features that make the effects of transcript information ambiguous. We find that in most cases \textsc{StockGNN} outperforms these methods. 





\subsection{Results on Different Labels}
\label{sec:acc_results}
We describe the results produced by our models on five major areas/sectors based on the number of transcripts.

\subsubsection{Experimental Settings}

We train the stock movement prediction methods on the transcripts till 2018. The transcripts in 2019 are used as test data. After removal of missing information, the train/test splits for all labels ($y_v$, $y_s$ and $y_{I,k}$) are shown in Table \ref{tab:train_test_both_label}.
We consider $k=5$ here and get a smaller subset (see Def. \ref{defn:ibl}) of the data.
The Doc2Vec method is applied to generate low-dimensional embeddings for the transcripts. We have generated embeddings with dimensions $100$, $200$, and $300$ and report the best results among these three. In most of the cases $300$-dimensional vector is the most effective one. The MLP classifier only uses one hidden layer with $32, 64, 128$ nodes for input features with dimensions $100$, $200$, and $300$ respectively and a sigmoid layer to generate the final classification. The \textsc{StockGNN} method uses an initial word vector of 300 dimensions as node (word) features and produces a $96$-dimensional vector as final graph-level embedding. We use the pre-trained GloVe\footnote{\url{http://nlp.stanford.edu/data/glove.6B.zip}} embeddings \cite{pennington2014glove} for these initial word (node) vectors. The other settings are described in more details in the Appendix (Section \ref{sec:extra_setting}).
\textbf{Performance Measures.} We measure the quality of our proposed methods by \textit{accuracy}, \textit{average (macro) precision} and \textit{average (macro) recall}.  We have also repeated all the experiments at least ten times and reported the mean performances.
 
\subsubsection{Results}
Our goal is to show the usefulness of the semantics of the transcripts in stock price movement prediction. We present the results in five major areas/sectors on the index based ($y_{I,5}$), shock based ($y_s$) and the value based ($y_v$) labels in Tables \ref{tab:results_strong_week}, \ref{tab:results_daily_shock} and \ref{tab:results_value_label} respectively.  Our \textsc{StockGNN} method produces the best quality in most of the cases. In particular, \textsc{StockGNN} produces $3\%, 4.1\%$ and $8.9\%$ more average (over all sectors) accuracy than \textsc{DEmlp}, \textsc{DEsvm},\textsc{DElogreg} respectively for index based label. 
Competitive results from the semantic feature based baselines indicate that the transcripts are indeed useful in the stock price movement prediction.



\textbf{Discussion:}  Prediction of stock price movement is a challenging task and often seen as a random walk process \cite{malkiel2003efficient}. As the investments in trading firms usually happen in large volumes---often in an automated fashion---lead to a significant amount of profit even with a slightly higher chance than random. Compared with a base rate or a random method, our method \textsc{StockGNN} shows significant improvement. Base rate assigns all the data as the label of the largest class whereas the random method assigns labels randomly. Both method have average recall of $0.5$. \textsc{StockGNN} achieves up to $23\%$, $22\%$ and $21\%$ more average recall than a base rate in the index, shock, and value based labels respectively. The performances of other baseline methods that use semantic features also validate that the earnings calls indeed have a correlation with stock price movements even over a period of week.

\begin{figure}[ht]
\vspace{-5mm}
    \centering
    \captionsetup[subfigure]{labelfont={normalsize,bf},textfont={normalsize,bf}}
    
    \subfloat[Avg. Recall ]{\includegraphics[width=0.22\textwidth]{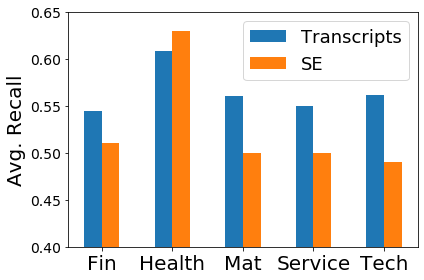}}
    \subfloat[Avg Precision ]{\includegraphics[width=0.22\textwidth]{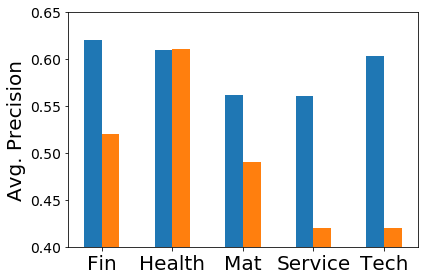}}
    
\vspace{-2mm}
    \caption{\textbf{Value Based Label ($y_v$) results:} (a) Average Recall and (b) Average Precision in five major areas/sectors. The features are constructed from semantic representations of the earnings call transcripts (Transcripts in blue) and the estimated and actual sales as well as EPS (SE in orange). The maximum standard deviations for Transcripts and SE  are $.003$ and $.002$ respectively. The semantic features from the transcripts produce more accurate results than SE in at least 80\% of the cases and also statistically significant (p-value $<0.01$). }
    \vspace{-4mm}
\label{fig:vs_SE}
\end{figure}
\subsection{ Baseline: Earnings per Share (EPS) \& Sales}
\label{sec:vs_SE}


Earnings per share (EPS) and sales are important quantitative factors that often are used to measure the performance of a company (Section \ref{sec:back_sales}). More specifically, if a company beats or misses the estimates in the sales and EPS for a given quarter, that usually results in higher and lower stock price movement respectively. Note that EPS (a real value) and earnings calls (a discussion) are different. Next, we show that the features from transcripts are more predictive of the stock price movement than the sales and EPS as features.

\subsubsection{Experimental Setting}

We have collected the estimated and actual sales and EPS corresponding to the earnings calls and used them as features to predict the value ($y_v$) as well as the index ($y_{I,5}$) based labels. We perform the classification tasks using a Multilayer Perceptron (MLP). In MLP we only use one hidden layer with $16$ nodes and a softmax layer to generate the final classification. The other classification methods have produced results of worse quality. 


\subsubsection{Results}
Figure \ref{fig:vs_SE} shows the results in five major areas for the value based label ($y_v$) prediction ( see Fig. \ref{fig:vs_SE_index} for the index based label $y_{I,5}$ in the Appendix).  We apply MLP in both scenarios---when the features are created from sales and EPS (denoted as SE) as well as when the features are semantic representations of the transcripts (denoted as Transcripts) produced by the Doc2Vec method (see Section \ref{sec:methods}). In most ($80\%$) of the cases, the semantic features from the transcripts produce more accurate results. In particular, the transcripts produce $15\%$ more AR and $42\%$ more AP in ``Tech'' and $10\%$ more AR and $33\%$ more AP in ``Service'' sectors. However, while transcripts produce better results in growing and evolving sectors such as ``Tech'', the values of the actual/estimated sales and EPS could be important for predicting stock price movement (e.g., ``Health''). Furthermore, Fig. \ref{fig:vis_vs_SE} presents a visualization of the difference in estimated and actual actual sales as well as EPS and their association with the value based labels (Sec. \ref{sec:sales_appendix} in the Appendix).




\section{ Related Work \& Background}




Sentiment analysis, tone and readability of the company related documents play a critical role in investments. Sentiment of the financial documents has been useful to predict market action \cite{kearney2014textual}. Even the clarity and readability of the financial documents help the investors in predicting the valuation whereas a complex and long text results in confusion and subsequent volatility \cite{loughran2014measuring}. Lee et al. \cite{lee2014importance} studied the reaction of the market in response to the reported 8-K documents. They found that the earning surprise is the most important feature in predicting price movements. Larcker and Zakolyukina \cite{larcker2012detecting} classified the earnings calls as "truthful" or "deceptive" based on subsequent statements from the company and shown that the language of deceptive executives exhibits more references to general knowledge, fewer non extreme positive emotions, and fewer references to shareholder value. 

In financial documents, the tone has also played an important role for stock market prediction \cite{loughran2011liability,henry2008investors,price2012earnings,davis2015effect,blau2015sophisticated}.
The tone in earnings press releases can influence investors \cite{henry2008investors}. Price et al. \cite{price2012earnings} found that the conference call linguistic tone is a significant predictor of abnormal returns and trading volumes. With analyses on the 10-Q and 10-K form, Feldman et al. \cite{feldman2010management} showed that market reactions in a short window around the SEC filing are significantly affected by the tone change, even after controlling for accruals and earnings surprises. Furthermore, the tone change can also help to predict the subsequent quarter’s earnings surprise. Moreover, the audio of the earnings call can also determine managerial emotional states \cite{mayew2012power}. Recently, Qin et al. \cite{qin2019you} used the audio and text of the earnings calls to predict stock volatility. 

New advancements in natural language
processing techniques allow new analysis of earning calls.  Our paper is inspired by the work of Ma et al. \cite{ma2020earnings} that addresses the problem of identifying the stock price movement via earnings calls. However, our classification task that measures weekly impact (Def. \ref{defn:ibl}) of an earnings call is novel and more practical. Our proposed graph neural network based method also outperforms the method in \cite{ma2020earnings} (more details are in Sec. \ref{sec:our_method}). Furthermore, our empirical analyses on a larger dataset are more elaborated, showing comparisons with external hard data such as sales and earnings per share (EPS) as well as the recommendations from the analysts that are missing in \cite{ma2020earnings}.

\textbf{Additional Details: }More details about Earnings per Share (EPS), sales, and analysts' ratings are provided in Sec. \ref{sec:back_sales} (Appendix).

\section{CONCLUSION}

 In this paper, we have studied the association between public companies’ earnings calls and stock price movements with a dataset of 97,478 transcripts from 6,300 different companies. We have demonstrated three novel findings using our method that exploits the semantic features of the earnings call transcripts. The semantic characteristics of transcripts are relatively more accurate to predict stock price movements. We have also established that the transcripts have more predictive power than traditional hard data such as actual and estimated values of sales and earnings per share. Interestingly, the recommendations from the analysts made prior to the earnings are unrelated to stock price movements after the earnings call whereas the sentiments in transcripts are valuable. 

\bibliographystyle{ACM-Reference-Format}
\bibliography{WWW22}


\begin{thebibliography}{37}


\ifx \showCODEN    \undefined \def \showCODEN     #1{\unskip}     \fi
\ifx \showDOI      \undefined \def \showDOI       #1{#1}\fi
\ifx \showISBNx    \undefined \def \showISBNx     #1{\unskip}     \fi
\ifx \showISBNxiii \undefined \def \showISBNxiii  #1{\unskip}     \fi
\ifx \showISSN     \undefined \def \showISSN      #1{\unskip}     \fi
\ifx \showLCCN     \undefined \def \showLCCN      #1{\unskip}     \fi
\ifx \shownote     \undefined \def \shownote      #1{#1}          \fi
\ifx \showarticletitle \undefined \def \showarticletitle #1{#1}   \fi
\ifx \showURL      \undefined \def \showURL       {\relax}        \fi
\providecommand\bibfield[2]{#2}
\providecommand\bibinfo[2]{#2}
\providecommand\natexlab[1]{#1}
\providecommand\showeprint[2][]{arXiv:#2}

\bibitem[\protect\citeauthoryear{Asquith, Mikhail, and Au}{Asquith
  et~al\mbox{.}}{2005}]%
        {asquith2005information}
\bibfield{author}{\bibinfo{person}{Paul Asquith}, \bibinfo{person}{Michael~B
  Mikhail}, {and} \bibinfo{person}{Andrea~S Au}.}
  \bibinfo{year}{2005}\natexlab{}.
\newblock \showarticletitle{Information content of equity analyst reports}.
\newblock \bibinfo{journal}{\emph{Journal of financial economics}}
  \bibinfo{volume}{75}, \bibinfo{number}{2} (\bibinfo{year}{2005}),
  \bibinfo{pages}{245--282}.
\newblock


\bibitem[\protect\citeauthoryear{Bagnoli, Kallapur, and Watts}{Bagnoli
  et~al\mbox{.}}{2001}]%
        {bagnoli2001top}
\bibfield{author}{\bibinfo{person}{Mark Bagnoli}, \bibinfo{person}{Sanjay
  Kallapur}, {and} \bibinfo{person}{Susan~G Watts}.}
  \bibinfo{year}{2001}\natexlab{}.
\newblock \showarticletitle{Top line and bottom line forecasts: a comparison of
  internet firms during and after the bubble}.
\newblock \bibinfo{journal}{\emph{Available at SSRN 274178}}
  (\bibinfo{year}{2001}).
\newblock


\bibitem[\protect\citeauthoryear{Barber, Lehavy, McNichols, and Trueman}{Barber
  et~al\mbox{.}}{2001}]%
        {barber2001can}
\bibfield{author}{\bibinfo{person}{Brad Barber}, \bibinfo{person}{Reuven
  Lehavy}, \bibinfo{person}{Maureen McNichols}, {and} \bibinfo{person}{Brett
  Trueman}.} \bibinfo{year}{2001}\natexlab{}.
\newblock \showarticletitle{Can investors profit from the prophets? Security
  analyst recommendations and stock returns}.
\newblock \bibinfo{journal}{\emph{The Journal of Finance}}
  \bibinfo{volume}{56}, \bibinfo{number}{2} (\bibinfo{year}{2001}),
  \bibinfo{pages}{531--563}.
\newblock


\bibitem[\protect\citeauthoryear{Barber, Lehavy, and Trueman}{Barber
  et~al\mbox{.}}{2007}]%
        {barber2007comparing}
\bibfield{author}{\bibinfo{person}{Brad~M Barber}, \bibinfo{person}{Reuven
  Lehavy}, {and} \bibinfo{person}{Brett Trueman}.}
  \bibinfo{year}{2007}\natexlab{}.
\newblock \showarticletitle{Comparing the stock recommendation performance of
  investment banks and independent research firms}.
\newblock \bibinfo{journal}{\emph{Journal of financial economics}}
  \bibinfo{volume}{85}, \bibinfo{number}{2} (\bibinfo{year}{2007}),
  \bibinfo{pages}{490--517}.
\newblock


\bibitem[\protect\citeauthoryear{Blau, DeLisle, and Price}{Blau
  et~al\mbox{.}}{2015}]%
        {blau2015sophisticated}
\bibfield{author}{\bibinfo{person}{Benjamin~M Blau}, \bibinfo{person}{Jared~R
  DeLisle}, {and} \bibinfo{person}{S~McKay Price}.}
  \bibinfo{year}{2015}\natexlab{}.
\newblock \showarticletitle{Do sophisticated investors interpret earnings
  conference call tone differently than investors at large? Evidence from short
  sales}.
\newblock \bibinfo{journal}{\emph{Journal of Corporate Finance}}
  \bibinfo{volume}{31} (\bibinfo{year}{2015}), \bibinfo{pages}{203--219}.
\newblock


\bibitem[\protect\citeauthoryear{Claeskens and Hjort}{Claeskens and
  Hjort}{2008}]%
        {RePEc:cup:cbooks:9780521852258}
\bibfield{author}{\bibinfo{person}{Gerda Claeskens} {and}
  \bibinfo{person}{Nils~Lid Hjort}.} \bibinfo{year}{2008}\natexlab{}.
\newblock \bibinfo{booktitle}{\emph{{Model Selection and Model Averaging}}}.
\newblock \bibinfo{publisher}{Cambridge University Press}.
\newblock


\bibitem[\protect\citeauthoryear{Davis, Ge, Matsumoto, and Zhang}{Davis
  et~al\mbox{.}}{2015}]%
        {davis2015effect}
\bibfield{author}{\bibinfo{person}{Angela~K Davis}, \bibinfo{person}{Weili Ge},
  \bibinfo{person}{Dawn Matsumoto}, {and} \bibinfo{person}{Jenny~Li Zhang}.}
  \bibinfo{year}{2015}\natexlab{}.
\newblock \showarticletitle{The effect of manager-specific optimism on the tone
  of earnings conference calls}.
\newblock \bibinfo{journal}{\emph{Review of Accounting Studies}}
  \bibinfo{volume}{20}, \bibinfo{number}{2} (\bibinfo{year}{2015}),
  \bibinfo{pages}{639--673}.
\newblock


\bibitem[\protect\citeauthoryear{Deng, Rangwala, and Ning}{Deng
  et~al\mbox{.}}{2019}]%
        {deng2019learning}
\bibfield{author}{\bibinfo{person}{Songgaojun Deng}, \bibinfo{person}{Huzefa
  Rangwala}, {and} \bibinfo{person}{Yue Ning}.}
  \bibinfo{year}{2019}\natexlab{}.
\newblock \showarticletitle{Learning dynamic context graphs for predicting
  social events}. In \bibinfo{booktitle}{\emph{KDD}}.
\newblock


\bibitem[\protect\citeauthoryear{Donders, Kouwenberg, and Vorst}{Donders
  et~al\mbox{.}}{2000}]%
        {donders2000options}
\bibfield{author}{\bibinfo{person}{WM Donders, Monique}, \bibinfo{person}{Roy
  Kouwenberg}, {and} \bibinfo{person}{CF Vorst, Ton}.}
  \bibinfo{year}{2000}\natexlab{}.
\newblock \showarticletitle{Options and earnings announcements: an empirical
  study of volatility, trading volume, open interest and liquidity}.
\newblock \bibinfo{journal}{\emph{European Financial Management}}
  \bibinfo{volume}{6}, \bibinfo{number}{2} (\bibinfo{year}{2000}),
  \bibinfo{pages}{149--171}.
\newblock


\bibitem[\protect\citeauthoryear{Ertimur, Livnat, and Martikainen}{Ertimur
  et~al\mbox{.}}{2003}]%
        {ertimur2003differential}
\bibfield{author}{\bibinfo{person}{Yonca Ertimur}, \bibinfo{person}{Joshua
  Livnat}, {and} \bibinfo{person}{Minna Martikainen}.}
  \bibinfo{year}{2003}\natexlab{}.
\newblock \showarticletitle{Differential market reactions to revenue and
  expense surprises}.
\newblock \bibinfo{journal}{\emph{Review of Accounting Studies}}
  \bibinfo{volume}{8}, \bibinfo{number}{2-3} (\bibinfo{year}{2003}),
  \bibinfo{pages}{185--211}.
\newblock


\bibitem[\protect\citeauthoryear{Feldman, Govindaraj, Livnat, and
  Segal}{Feldman et~al\mbox{.}}{2010}]%
        {feldman2010management}
\bibfield{author}{\bibinfo{person}{Ronen Feldman}, \bibinfo{person}{Suresh
  Govindaraj}, \bibinfo{person}{Joshua Livnat}, {and} \bibinfo{person}{Benjamin
  Segal}.} \bibinfo{year}{2010}\natexlab{}.
\newblock \showarticletitle{Management’s tone change, post earnings
  announcement drift and accruals}.
\newblock \bibinfo{journal}{\emph{Review of Accounting Studies}}
  \bibinfo{volume}{15}, \bibinfo{number}{4} (\bibinfo{year}{2010}),
  \bibinfo{pages}{915--953}.
\newblock


\bibitem[\protect\citeauthoryear{Francis and Soffer}{Francis and
  Soffer}{1997}]%
        {francis1997relative}
\bibfield{author}{\bibinfo{person}{Jennifer Francis} {and}
  \bibinfo{person}{Leonard Soffer}.} \bibinfo{year}{1997}\natexlab{}.
\newblock \showarticletitle{The relative informativeness of analysts' stock
  recommendations and earnings forecast revisions}.
\newblock \bibinfo{journal}{\emph{Journal of Accounting Research}}
  \bibinfo{volume}{35}, \bibinfo{number}{2} (\bibinfo{year}{1997}),
  \bibinfo{pages}{193--211}.
\newblock


\bibitem[\protect\citeauthoryear{Henry}{Henry}{2008}]%
        {henry2008investors}
\bibfield{author}{\bibinfo{person}{Elaine Henry}.}
  \bibinfo{year}{2008}\natexlab{}.
\newblock \showarticletitle{Are investors influenced by how earnings press
  releases are written?}
\newblock \bibinfo{journal}{\emph{The Journal of Business Communication
  (1973)}} \bibinfo{volume}{45}, \bibinfo{number}{4} (\bibinfo{year}{2008}),
  \bibinfo{pages}{363--407}.
\newblock


\bibitem[\protect\citeauthoryear{Hu, Liu, Bian, Liu, and Liu}{Hu
  et~al\mbox{.}}{2018}]%
        {hu2018listening}
\bibfield{author}{\bibinfo{person}{Ziniu Hu}, \bibinfo{person}{Weiqing Liu},
  \bibinfo{person}{Jiang Bian}, \bibinfo{person}{Xuanzhe Liu}, {and}
  \bibinfo{person}{Tie-Yan Liu}.} \bibinfo{year}{2018}\natexlab{}.
\newblock \showarticletitle{Listening to chaotic whispers: A deep learning
  framework for news-oriented stock trend prediction}. In
  \bibinfo{booktitle}{\emph{WSDM}}. \bibinfo{pages}{261--269}.
\newblock


\bibitem[\protect\citeauthoryear{Huang, Ma, Li, Zhang, and Wang}{Huang
  et~al\mbox{.}}{2019}]%
        {huang2019text}
\bibfield{author}{\bibinfo{person}{Lianzhe Huang}, \bibinfo{person}{Dehong Ma},
  \bibinfo{person}{Sujian Li}, \bibinfo{person}{Xiaodong Zhang}, {and}
  \bibinfo{person}{Houfeng Wang}.} \bibinfo{year}{2019}\natexlab{}.
\newblock \showarticletitle{Text level graph neural network for text
  classification}. In \bibinfo{booktitle}{\emph{ACL}}.
\newblock


\bibitem[\protect\citeauthoryear{Ivkovi{\'c} and Jegadeesh}{Ivkovi{\'c} and
  Jegadeesh}{2004}]%
        {ivkovic2004timing}
\bibfield{author}{\bibinfo{person}{Zoran Ivkovi{\'c}} {and}
  \bibinfo{person}{Narasimhan Jegadeesh}.} \bibinfo{year}{2004}\natexlab{}.
\newblock \showarticletitle{The timing and value of forecast and recommendation
  revisions}.
\newblock \bibinfo{journal}{\emph{Journal of Financial Economics}}
  \bibinfo{volume}{73}, \bibinfo{number}{3} (\bibinfo{year}{2004}),
  \bibinfo{pages}{433--463}.
\newblock


\bibitem[\protect\citeauthoryear{Kearney and Liu}{Kearney and Liu}{2014}]%
        {kearney2014textual}
\bibfield{author}{\bibinfo{person}{Colm Kearney} {and} \bibinfo{person}{Sha
  Liu}.} \bibinfo{year}{2014}\natexlab{}.
\newblock \showarticletitle{Textual sentiment in finance: A survey of methods
  and models}.
\newblock \bibinfo{journal}{\emph{International Review of Financial Analysis}}
  \bibinfo{volume}{33} (\bibinfo{year}{2014}), \bibinfo{pages}{171--185}.
\newblock


\bibitem[\protect\citeauthoryear{Kipf and Welling}{Kipf and Welling}{2016}]%
        {kipf2016semi}
\bibfield{author}{\bibinfo{person}{Thomas~N Kipf} {and} \bibinfo{person}{Max
  Welling}.} \bibinfo{year}{2016}\natexlab{}.
\newblock \showarticletitle{Semi-Supervised Classification with Graph
  Convolutional Networks}.
\newblock \bibinfo{journal}{\emph{arXiv preprint arXiv:1609.02907}}
  (\bibinfo{year}{2016}).
\newblock


\bibitem[\protect\citeauthoryear{Kosan, Silva, Medya, Uzzi, and Singh}{Kosan
  et~al\mbox{.}}{2021}]%
        {kosan2021event}
\bibfield{author}{\bibinfo{person}{Mert Kosan}, \bibinfo{person}{Arlei Silva},
  \bibinfo{person}{Sourav Medya}, \bibinfo{person}{Brian Uzzi}, {and}
  \bibinfo{person}{Ambuj Singh}.} \bibinfo{year}{2021}\natexlab{}.
\newblock \showarticletitle{Event Detection on Dynamic Graphs}.
\newblock \bibinfo{journal}{\emph{arXiv preprint arXiv:2110.12148}}
  (\bibinfo{year}{2021}).
\newblock


\bibitem[\protect\citeauthoryear{Larcker and Zakolyukina}{Larcker and
  Zakolyukina}{2012}]%
        {larcker2012detecting}
\bibfield{author}{\bibinfo{person}{David~F Larcker} {and}
  \bibinfo{person}{Anastasia~A Zakolyukina}.} \bibinfo{year}{2012}\natexlab{}.
\newblock \showarticletitle{Detecting deceptive discussions in conference
  calls}.
\newblock \bibinfo{journal}{\emph{Journal of Accounting Research}}
  \bibinfo{volume}{50}, \bibinfo{number}{2} (\bibinfo{year}{2012}),
  \bibinfo{pages}{495--540}.
\newblock


\bibitem[\protect\citeauthoryear{Le and Mikolov}{Le and Mikolov}{2014}]%
        {le2014distributed}
\bibfield{author}{\bibinfo{person}{Quoc Le} {and} \bibinfo{person}{Tomas
  Mikolov}.} \bibinfo{year}{2014}\natexlab{}.
\newblock \showarticletitle{Distributed representations of sentences and
  documents}. In \bibinfo{booktitle}{\emph{ICML}}. \bibinfo{pages}{1188--1196}.
\newblock


\bibitem[\protect\citeauthoryear{Lee, Surdeanu, MacCartney, and Jurafsky}{Lee
  et~al\mbox{.}}{2014}]%
        {lee2014importance}
\bibfield{author}{\bibinfo{person}{Heeyoung Lee}, \bibinfo{person}{Mihai
  Surdeanu}, \bibinfo{person}{Bill MacCartney}, {and} \bibinfo{person}{Dan
  Jurafsky}.} \bibinfo{year}{2014}\natexlab{}.
\newblock \showarticletitle{On the Importance of Text Analysis for Stock Price
  Prediction.}. In \bibinfo{booktitle}{\emph{LREC}},
  Vol.~\bibinfo{volume}{2014}. \bibinfo{pages}{1170--1175}.
\newblock


\bibitem[\protect\citeauthoryear{Lev}{Lev}{1989}]%
        {lev1989usefulness}
\bibfield{author}{\bibinfo{person}{Baruch Lev}.}
  \bibinfo{year}{1989}\natexlab{}.
\newblock \showarticletitle{On the usefulness of earnings and earnings
  research: Lessons and directions from two decades of empirical research}.
\newblock \bibinfo{journal}{\emph{Journal of accounting research}}
  \bibinfo{volume}{27} (\bibinfo{year}{1989}), \bibinfo{pages}{153--192}.
\newblock


\bibitem[\protect\citeauthoryear{Li, Tarlow, Brockschmidt, and Zemel}{Li
  et~al\mbox{.}}{2015}]%
        {li2015gated}
\bibfield{author}{\bibinfo{person}{Yujia Li}, \bibinfo{person}{Daniel Tarlow},
  \bibinfo{person}{Marc Brockschmidt}, {and} \bibinfo{person}{Richard Zemel}.}
  \bibinfo{year}{2015}\natexlab{}.
\newblock \showarticletitle{Gated graph sequence neural networks}.
\newblock \bibinfo{journal}{\emph{arXiv preprint arXiv:1511.05493}}
  (\bibinfo{year}{2015}).
\newblock


\bibitem[\protect\citeauthoryear{Loh and Stulz}{Loh and Stulz}{2011}]%
        {loh2011analyst}
\bibfield{author}{\bibinfo{person}{Roger~K Loh} {and}
  \bibinfo{person}{Ren{\'e}~M Stulz}.} \bibinfo{year}{2011}\natexlab{}.
\newblock \showarticletitle{When are analyst recommendation changes
  influential?}
\newblock \bibinfo{journal}{\emph{The review of financial studies}}
  \bibinfo{volume}{24}, \bibinfo{number}{2} (\bibinfo{year}{2011}),
  \bibinfo{pages}{593--627}.
\newblock


\bibitem[\protect\citeauthoryear{Loughran and McDonald}{Loughran and
  McDonald}{2011}]%
        {loughran2011liability}
\bibfield{author}{\bibinfo{person}{Tim Loughran} {and} \bibinfo{person}{Bill
  McDonald}.} \bibinfo{year}{2011}\natexlab{}.
\newblock \showarticletitle{When is a liability not a liability? Textual
  analysis, dictionaries, and 10-Ks}.
\newblock \bibinfo{journal}{\emph{The Journal of Finance}}
  \bibinfo{volume}{66}, \bibinfo{number}{1} (\bibinfo{year}{2011}),
  \bibinfo{pages}{35--65}.
\newblock


\bibitem[\protect\citeauthoryear{Loughran and McDonald}{Loughran and
  McDonald}{2014}]%
        {loughran2014measuring}
\bibfield{author}{\bibinfo{person}{Tim Loughran} {and} \bibinfo{person}{Bill
  McDonald}.} \bibinfo{year}{2014}\natexlab{}.
\newblock \showarticletitle{Measuring readability in financial disclosures}.
\newblock \bibinfo{journal}{\emph{The Journal of Finance}}
  \bibinfo{volume}{69}, \bibinfo{number}{4} (\bibinfo{year}{2014}),
  \bibinfo{pages}{1643--1671}.
\newblock


\bibitem[\protect\citeauthoryear{Ma, Bang, Wang, and Liu}{Ma
  et~al\mbox{.}}{2020}]%
        {ma2020earnings}
\bibfield{author}{\bibinfo{person}{Zhiqiang Ma}, \bibinfo{person}{Grace Bang},
  \bibinfo{person}{Chong Wang}, {and} \bibinfo{person}{Xiaomo Liu}.}
  \bibinfo{year}{2020}\natexlab{}.
\newblock \showarticletitle{Towards Earnings Call and Stock Price Movement}. In
  \bibinfo{booktitle}{\emph{SIGKDD MLF Workshop}}.
\newblock


\bibitem[\protect\citeauthoryear{Malkiel}{Malkiel}{2003}]%
        {malkiel2003efficient}
\bibfield{author}{\bibinfo{person}{Burton~G Malkiel}.}
  \bibinfo{year}{2003}\natexlab{}.
\newblock \showarticletitle{The efficient market hypothesis and its critics}.
\newblock \bibinfo{journal}{\emph{Journal of economic perspectives}}
  \bibinfo{volume}{17}, \bibinfo{number}{1} (\bibinfo{year}{2003}),
  \bibinfo{pages}{59--82}.
\newblock


\bibitem[\protect\citeauthoryear{Mayew and Venkatachalam}{Mayew and
  Venkatachalam}{2012}]%
        {mayew2012power}
\bibfield{author}{\bibinfo{person}{William~J Mayew} {and}
  \bibinfo{person}{Mohan Venkatachalam}.} \bibinfo{year}{2012}\natexlab{}.
\newblock \showarticletitle{The power of voice: Managerial affective states and
  future firm performance}.
\newblock \bibinfo{journal}{\emph{The Journal of Finance}}
  \bibinfo{volume}{67}, \bibinfo{number}{1} (\bibinfo{year}{2012}),
  \bibinfo{pages}{1--43}.
\newblock


\bibitem[\protect\citeauthoryear{Pennington, Socher, and Manning}{Pennington
  et~al\mbox{.}}{2014}]%
        {pennington2014glove}
\bibfield{author}{\bibinfo{person}{Jeffrey Pennington},
  \bibinfo{person}{Richard Socher}, {and} \bibinfo{person}{Christopher~D
  Manning}.} \bibinfo{year}{2014}\natexlab{}.
\newblock \showarticletitle{Glove: Global vectors for word representation}. In
  \bibinfo{booktitle}{\emph{EMNLP}}. \bibinfo{pages}{1532--1543}.
\newblock


\bibitem[\protect\citeauthoryear{Price, Doran, Peterson, and Bliss}{Price
  et~al\mbox{.}}{2012}]%
        {price2012earnings}
\bibfield{author}{\bibinfo{person}{S~McKay Price}, \bibinfo{person}{James~S
  Doran}, \bibinfo{person}{David~R Peterson}, {and} \bibinfo{person}{Barbara~A
  Bliss}.} \bibinfo{year}{2012}\natexlab{}.
\newblock \showarticletitle{Earnings conference calls and stock returns: The
  incremental informativeness of textual tone}.
\newblock \bibinfo{journal}{\emph{Journal of Banking \& Finance}}
  \bibinfo{volume}{36}, \bibinfo{number}{4} (\bibinfo{year}{2012}),
  \bibinfo{pages}{992--1011}.
\newblock


\bibitem[\protect\citeauthoryear{Qin and Yang}{Qin and Yang}{2019}]%
        {qin2019you}
\bibfield{author}{\bibinfo{person}{Yu Qin} {and} \bibinfo{person}{Yi Yang}.}
  \bibinfo{year}{2019}\natexlab{}.
\newblock \showarticletitle{What you say and how you say it matters: Predicting
  stock volatility using verbal and vocal cues}. In
  \bibinfo{booktitle}{\emph{ACL}}. \bibinfo{pages}{390--401}.
\newblock


\bibitem[\protect\citeauthoryear{Stickel}{Stickel}{1995}]%
        {stickel1995anatomy}
\bibfield{author}{\bibinfo{person}{Scott~E Stickel}.}
  \bibinfo{year}{1995}\natexlab{}.
\newblock \showarticletitle{The anatomy of the performance of buy and sell
  recommendations}.
\newblock \bibinfo{journal}{\emph{Financial Analysts Journal}}
  \bibinfo{volume}{51}, \bibinfo{number}{5} (\bibinfo{year}{1995}),
  \bibinfo{pages}{25--39}.
\newblock


\bibitem[\protect\citeauthoryear{Wasserstein and Lazar}{Wasserstein and
  Lazar}{2016}]%
        {wasserstein2016asa}
\bibfield{author}{\bibinfo{person}{Ronald~L Wasserstein} {and}
  \bibinfo{person}{Nicole~A Lazar}.} \bibinfo{year}{2016}\natexlab{}.
\newblock \bibinfo{title}{The ASA statement on p-values: context, process, and
  purpose}.
\newblock
\newblock


\bibitem[\protect\citeauthoryear{Xu and Cohen}{Xu and Cohen}{2018}]%
        {xu2018stock}
\bibfield{author}{\bibinfo{person}{Yumo Xu} {and} \bibinfo{person}{Shay~B
  Cohen}.} \bibinfo{year}{2018}\natexlab{}.
\newblock \showarticletitle{Stock movement prediction from tweets and
  historical prices}. In \bibinfo{booktitle}{\emph{ACL}}.
  \bibinfo{pages}{1970--1979}.
\newblock


\bibitem[\protect\citeauthoryear{Zhang, Yu, Cui, Wu, Wen, and Wang}{Zhang
  et~al\mbox{.}}{2020}]%
        {zhang2020every}
\bibfield{author}{\bibinfo{person}{Yufeng Zhang}, \bibinfo{person}{Xueli Yu},
  \bibinfo{person}{Zeyu Cui}, \bibinfo{person}{Shu Wu},
  \bibinfo{person}{Zhongzhen Wen}, {and} \bibinfo{person}{Liang Wang}.}
  \bibinfo{year}{2020}\natexlab{}.
\newblock \showarticletitle{Every Document Owns Its Structure: Inductive Text
  Classification via Graph Neural Networks}. In
  \bibinfo{booktitle}{\emph{ACL}}.
\newblock


\end{thebibliography}

\clearpage

\section{Appendix}

\subsection{Sales, Earnings \& Analysts' Ratings}

\label{sec:back_sales}


\textbf{Earnings per Share (EPS) and Sales:} EPS and sales are two important quantitative factors that are often used to evaluate the performance of a company. The EPS indicates whether the company is currently profitable and can be self-sustained. The investors are often rewarded based on the EPS of a company. Moreover, the stock price is usually a multiplier (i.e., price earning ratio) of the EPS. The EPS often plays an important role in the stock movement especially for the established companies with small growth \cite{bagnoli2001top}. Note that the EPS (a real value) and earnings calls (a discussion) are different. On the other hand, sales is indicative of the market share of a company and an important measure especially for growth companies \cite{bagnoli2001top}. Sales might influence investors to overlook or ignore the earnings while having the promise of future growth \cite{lev1989usefulness}. 



Sales and EPS are important quantitative factors that often affect the perception of the investors about corresponding companies. Usually before the earnings call happens, the market forms a general consensus (estimates) on the EPS and sales. If a company beats (or misses) the estimates in the sales and EPS for a given quarter, that usually results in higher (or lower) stock price movement. Past research has shown that earning surprise due to surprise in sales has more effect on the price than due to cutting costs \cite{ertimur2003differential}. In this study, we demonstrate that the transcripts have more predictive power than  actual and estimated values of sales and EPS.

\textbf{Analysts' Ratings:}
Analysts are the financial experts who often post their opinions about companies and play a key role in impacting or modifying the perceptions of the investors about the future prospects of a business \cite{barber2001can,francis1997relative,asquith2005information}. Past research has demonstrated that analysts from popular banking and investment firms can generate significant stock price movements \cite{loh2011analyst,stickel1995anatomy,barber2007comparing}. Analysts' recommendations become even more powerful when they are accompanied by forecasts of the earnings of the company \cite{loh2011analyst}. Moreover, the timing of these recommendations with respect to the earnings announcements affects their impact \cite{ivkovic2004timing}. In this work, we analyze the impact of the recommendations by the analysts in stock price movements locally after the occurrence of the earnings call. 

\subsection{Additional Regression Results}
In Table~\ref{tab:shock_label_prediction} we present the same analyses as in Table~\ref{tab:daily_movement_prediction} for the shock based label $y_s$ as a dependent variable. The analyses produce similar results as in the case of value based label $y_v$ .

\begin{table}[]
\footnotesize
\begin{tabular}{llllll}
Variables         & Model 1 & Model 2                                                    & Model 3                                                   & Model 4                                                   & Model 5                                                      \\
Positive Emotion  &         & \begin{tabular}[c]{@{}l@{}}0.32***\\ (0.027)\end{tabular}  & \begin{tabular}[c]{@{}l@{}}0.31***\\ (0.027)\end{tabular} & \begin{tabular}[c]{@{}l@{}}0.33***\\ (0.028)\end{tabular} & \begin{tabular}[c]{@{}l@{}}0.42***\\ (0.036)\end{tabular}    \\
Negative Emotion  &         & \begin{tabular}[c]{@{}l@{}}-0.80***\\ (0.069)\end{tabular} & \begin{tabular}[c]{@{}l@{}}-0.29*\\ (0.11)\end{tabular}   & \begin{tabular}[c]{@{}l@{}}-0.31**\\ (0.11)\end{tabular}  & \begin{tabular}[c]{@{}l@{}}-0.45***\\ (0.15)\end{tabular}    \\
Anxiety Score     &         &                                                            & \begin{tabular}[c]{@{}l@{}}-0.43*\\ (0.22)\end{tabular}   & \begin{tabular}[c]{@{}l@{}}-0.38\\ (0.22)\end{tabular}    & \begin{tabular}[c]{@{}l@{}}-0.12\\ (0.29)\end{tabular}       \\
Anger Score       &         &                                                            & \begin{tabular}[c]{@{}l@{}}0.042\\ (0.25)\end{tabular}    & \begin{tabular}[c]{@{}l@{}}0.080\\ (0.25)\end{tabular}    & \begin{tabular}[c]{@{}l@{}}0.17\\ (0.31)\end{tabular}        \\
Sad Score         &         &                                                            & \begin{tabular}[c]{@{}l@{}}-1.24***\\ (0.19)\end{tabular} & \begin{tabular}[c]{@{}l@{}}-1.26***\\ (0.19)\end{tabular} & \begin{tabular}[c]{@{}l@{}}-1.78***\\ (0.26)\end{tabular}    \\
Certainty Score   &         &                                                            & \begin{tabular}[c]{@{}l@{}}0.041\\ (0.068)\end{tabular}   & \begin{tabular}[c]{@{}l@{}}0.083\\ (0.074)\end{tabular}   & \begin{tabular}[c]{@{}l@{}}0.18\\ (0.098)\end{tabular}       \\
Cognitive Score   &         &                                                            &                                                           & \begin{tabular}[c]{@{}l@{}}-0.032*\\ (0.015)\end{tabular} & \begin{tabular}[c]{@{}l@{}}-0.058*\\ (0.021)\end{tabular}    \\
Insight Score     &         &                                                            &                                                           & \begin{tabular}[c]{@{}l@{}}-0.060\\ (0.043)\end{tabular}  & \begin{tabular}[c]{@{}l@{}}-0.11\\ (0.059)\end{tabular}      \\
Causation Score   &         &                                                            &                                                           & \begin{tabular}[c]{@{}l@{}}-0.11*\\ (0.042)\end{tabular}  & \begin{tabular}[c]{@{}l@{}}-0.00077\\ (0.053)\end{tabular}   \\
Discrepancy Score &         &                                                            &                                                           & \begin{tabular}[c]{@{}l@{}}0.19*\\ (0.063)\end{tabular}   & \begin{tabular}[c]{@{}l@{}}0.19*\\ (0.082)\end{tabular}      \\
Actual Sales      &         &                                                            &                                                           &                                                           & \begin{tabular}[c]{@{}l@{}}0.00052\\ (0.00028)\end{tabular}  \\
Estimated Sales   &         &                                                            &                                                           &                                                           & \begin{tabular}[c]{@{}l@{}}-0.00053\\ (0.00028)\end{tabular} \\
Actual EPS        &         &                                                            &                                                           &                                                           & \begin{tabular}[c]{@{}l@{}}0.042\\ (0.036)\end{tabular}      \\
Estimated EPS     &         &                                                            &                                                           &                                                           & \begin{tabular}[c]{@{}l@{}}-0.033\\ (0.029)\end{tabular}     \\
Sector            & YES     & YES                                                        & YES                                                       & YES                                                       & YES                                                          \\
Year              & YES     & YES                                                        & YES                                                       & YES                                                       & YES                                                          \\
Observations      & 20,020  & 19,965                                                     & 19,965                                                    & 19,965                                                    & 12,946                                    \\
BIC               & 27,887  & 27,463                                                     & 27,437                                                    & 27,445                                                    & 17,695                                                      
\end{tabular}
\caption{Regression of Shock Based Label Function (SBL) $y_s(T^c_d)$ on Text Features.}
\label{tab:shock_label_prediction}
\vspace{-2mm}
\end{table}

\subsubsection{Analysts' Recommendations}
\label{sec:extra_analyst}

 With the same set of independent variables as in Table~\ref{reg_mar_3_1_value}, we present the regression results for the for $y_{I,k}$ (Index Based Label Function) when $k=3$ in Tables~\ref{reg_mar_3_1_index} ($k=4, 5$ produce similar results). Different from results in Table~\ref{reg_mar_3_1_value} for $y_v$, we find that only the positive emotion in the transcripts remains significantly (low $p$-value) indicative of the label $y_{I,3}$ when other control variables are added into the model incrementally. However, some results are consistent with the case of $y_v$. Adding more control variables does not improve the fitness of the model. BIC in Model 2 is larger (worse) than that of Model 1, which suggests that prior analysts' recommendations do not provide additional information to the model.  

\subsection{Experimental Settings of Section \ref{sec:acc_results}}
\label{sec:extra_setting}
The Doc2Vec method generates low-dimensional embeddings for the transcripts and uses standard settings as follows: ``distributed memory" (PV-DM), the initial learning rate as $.001$, and the negative sample as $5$. We ignore all words with total frequency lower than $2$. In the \textsc{StockGNN} method we use the window size of $3$ to build the graph from a document. The number of iterations (steps, $k$) that the gated graph neural network (GNN) uses inside the method is $2$. We use the batch size of $128$ in all the experiments. 

\begin{table}[t]
\footnotesize
\begin{tabular}{|l|c c c c c|}
\hline
Variables    & Model   1 & Model   2 & Model   3 & Model   4 & Model   5 \\ \hline 
positive     & 0.10***   & 0.10***   & 0.10***   & 0.10***   & 0.12**    \\ 
             & (0.03)    & (0.03)    & (0.03)    & (0.03)    & (0.04)    \\ \hline
negative     & -0.10     & -0.10     & -0.14     & -0.14     & -0.07     \\ 
             & (0.12)    & (0.12)    & (0.13)    & (0.13)    & (0.18)    \\ \hline
anxiety      & 0.28      & 0.28      & 0.21      & 0.19      & 0.09      \\ 
             & (0.22)    & (0.22)    & (0.24)    & (0.24)    & (0.34)    \\ \hline
anger        & 0.26      & 0.26      & 0.42      & 0.41      & 0.54      \\ 
             & (0.24)    & (0.24)    & (0.24)    & (0.24)    & (0.36)    \\ \hline
sad          & -0.02     & -0.03     & 0.03      & 0.03      & -0.06     \\ 
             & (0.18)    & (0.18)    & (0.19)    & (0.19)    & (0.29)    \\ \hline
certain      & 0.12      & 0.12      & 0.11      & 0.11      & 0.08      \\ 
             & (0.07)    & (0.07)    & (0.07)    & (0.07)    & (0.10)    \\ \hline
$MAR_{1m}$     &           & YES       & YES       & YES       & YES       \\ \hline
Sector     &           &           & YES       & YES       & YES       \\ \hline
Year         &           &           &           & YES       & YES       \\ \hline
$MAR_{5d}$     &           &           &           &           & YES       \\ \hline
Observations & 18,545    & 18,545    & 18,545    & 18,545    & 9,097     \\ 
BIC          & 25733     & 25771     & 25836     & 25893     & 12823     \\ \hline
\end{tabular}
\caption{Index Based Label Function ($y_{I,3}$) results on Analysts' Recommendations within 1 month ($MAR_{1m}$) prior and 5 days ($MAR_{5d}$) after the earnings call date and sentiment of the transcripts.} 
\vspace{-4mm}
\label{reg_mar_3_1_index}
\end{table}

\begin{figure*}[ht]
 \vspace{-3mm}
    \centering
    \captionsetup[subfigure]{labelfont={normalsize,bf},textfont={normalsize,bf}}
    
    \includegraphics[width=0.8\textwidth]{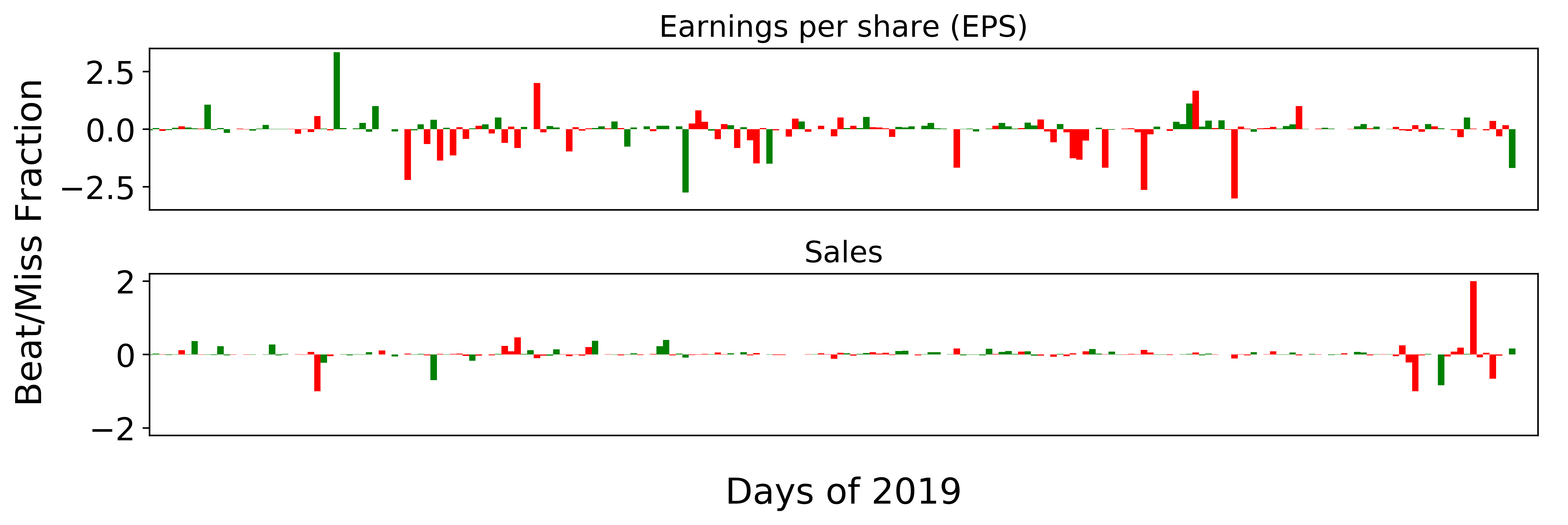}
    \vspace{-4mm}

    \caption{\textbf{Value Based Label ($y_{v}$) results:} It presents a visualization of the difference in estimated and actual sales as well as earnings per share (EPS) and their association with the labels. The X-axis shows a total of 212 days which has at least one occurrence of earnings call from any company in 2019. The Y-axis shows the factor by which the actual value beats (positive and upward) or misses (negative and downward) the estimated value. Some of the high positive (or negative) values of these factors predict the class 1 (or 0) labels, denoted by upward green (or downward red), accurately. However, there are several mis-classifications in the form of upward red and downward green in both cases (Sales and EPS).   } 
\label{fig:vis_vs_SE}
\end{figure*}

\begin{figure}[h]
\vspace{-4mm}
    \centering
    \captionsetup[subfigure]{labelfont={normalsize,bf},textfont={normalsize,bf}}
    
    \subfloat[Avg. Recall ]{\includegraphics[width=0.24\textwidth]{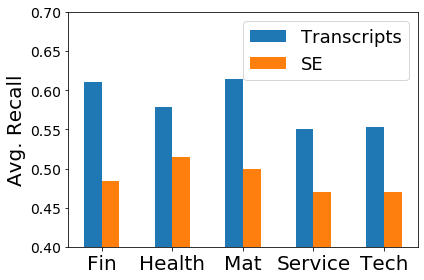}}
    \subfloat[Avg Precision ]{\includegraphics[width=0.24\textwidth]{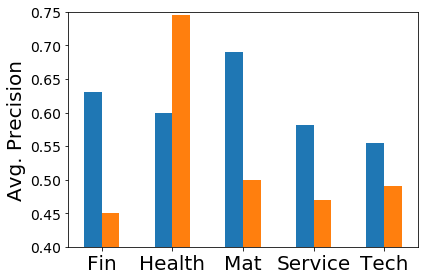}}
    
\vspace{-2mm}
    \caption{\textbf{IBL ($y_{I,5}$) results:} (a) Average Recall and (b) Average Precision in five major areas/sectors. The features are constructed from the earnings calls transcripts (Transcripts, denoted by blue) and the estimated and actual sales as well as EPS (SE, denoted by orange). The semantic features from the transcripts produce more accurate results than SE in at least 80\% of the cases.  } 
\label{fig:vs_SE_index}
\end{figure}

\subsection{Results: Earnings per Share and Sales}
\label{sec:sales_appendix}

\textbf{Visualizations: }In Figure \ref{fig:vs_SE} (Section \ref{sec:vs_SE}), we have already shown that the features with estimated/actual sales and EPS as features are inferior than semantics of transcripts in predicting stock price movements for the value based labels ($y_v$).  Figure \ref{fig:vis_vs_SE} presents a visualization of the differences in estimated and actual sales as well as earnings per share (EPS) and their association with the value based labels ($y_v$). The X-axis shows a total of 212 days on which there is at least one occurrence of earnings call in 2019. If there are multiple earnings calls on the same date, we randomly choose one. The Y-axis shows the factor by which the actual value beats (positive) or misses (negative) the estimated value. For instance, $0.5$ would mean the actual value beat the estimated one by $50\%$. The figure demonstrates that some of the high positive (or negative) values of these factors predict the class 1 (or 0) labels, denoted by upward green (or downward red), accurately. However, there are several misclassifications in the form of upward red and downward green in both cases (sales and EPS). These imply that beating (missing) the estimated value is not always associated with upward (downward) stock movement for both sales and EPS.  

\textbf{Results on index based label $y_{I,5}$: }We present additional results for the index based label $y_{I,5}$ in five major sectors in Figure \ref{fig:vs_SE_index}. The classification results are produced by a Multilayer Perceptron (MLP) in two scenarios---when the features are created from sales and EPS (denoted as SE) as well as when the features are semantic representations of the transcripts (denoted as Transcripts) produced by the Doc2Vec method (see Section \ref{sec:methods}). In 80\% of the cases, the semantic features from the transcripts produce more accurate results. In particular, the transcripts produce up to $21\%$ more AR and $40\%$ more AP in finance (``Fin") sector. 

\subsection{Effect of Sentiments and Semantics}
\label{sec:senti_sem}
\begin{figure}[t]
\vspace{-3mm}
    \centering
    \captionsetup[subfigure]{labelfont={normalsize,bf},textfont={normalsize,bf}}
    \subfloat[Avg. Recall]{\includegraphics[width=0.22\textwidth]{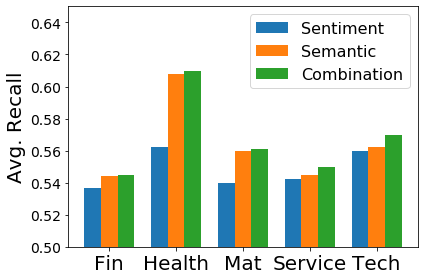}}
    \subfloat[ Avg. Precision]{\includegraphics[width=0.22\textwidth]{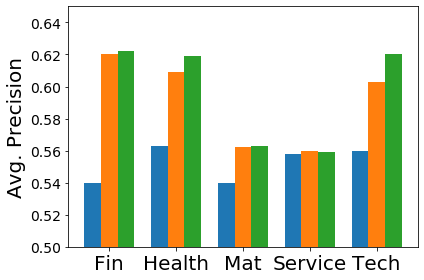}}
    \vspace{-2mm}
    \caption{\textbf{Value Based Label ($y_v$) results:} (a) Average Recall and (b) Average Precision of predicting the labels in five major sectors. The features are constructed from sentiments (blue), semantic representations of the transcripts (orange), and combination of them (green). The results show that though the semantic features are relatively the most predictive features, sentiments often play a critical role.} 
\label{fig:model_sentiment_VBL}
\end{figure}

The previous results indicate that the sentiment variables from transcripts might be useful for stock price movement prediction. We further use those six sentiment variables as features and compare the predictions with the ones that use the semantic features produced by our \textsc{StockGNN} method (see Section \ref{sec:methods}). To investigate whether these two features capture different information, we also present the results of the combination of both as features. Figure \ref{fig:model_sentiment_VBL} presents the classification results produced by a Multilayer Perceptron (MLP). As performance measures, we show average precision and average recall. Though the semantic features are the most useful ones in all of the five sectors, sentiment in transcripts is also useful for prediction. For instance, in the ``Service'' sector, the sentiment features produce competitive results compared to the semantic ones. 

\end{document}